\documentclass[aps,10pt,pra,twocolumn,showpacs,showkeys,groupedaddress]{revtex4-1}
 
\usepackage{amsmath}
\usepackage{graphicx}
\usepackage{bm}
\usepackage{xcolor}
\usepackage{comment}
\usepackage[normalem]{ulem}
\usepackage{enumitem}
\usepackage{natbib}
 
\begin{document}

\newcommand{\be}{\begin{equation}}
\newcommand{\ee}{\end{equation}}
 
\title{Generalized wave-packet model for studying coherence of matter-wave interferometers}
 
\author{Y. Japha}    
\email[Email:]{japhay@bgu.ac.il}
\affiliation{Department of Physics, Ben-Gurion University of the Negev, Be'er Sheva 84105, Israel}
 
\begin{abstract}
 
We present a generalized model for the evolution of atomic wave-packets in matter-wave interferometers. This method provides an efficient tool for analyzing the performance of atomic interferometers using atom clouds prepared in a trap as a Bose-Einstein condensate (BEC) or as a thermal ensemble. Predictions of the model for dynamic properties such as wave-packet size and phase are in excellent agreement with explicit numerical solutions of the non-linear Gross-Pitaevskii equations and enable fast calculations of interferometric performance in regimes where full numerical solutions become impractical. 
As a starting point, the static Thomas-Fermi (TF) approximation for a BEC in a harmonic trap is generalized to the whole range of atom-atom interaction strengths: from  non-interacting atoms (low densities) to the standard TF limit (high atomic densities, as long as the condensate approximation still holds). 
In particular, this generalization allows a good estimation of atomic cloud properties along the transition from a three-dimensional to a quasi-one-dimensional BEC in an elongated trap. 
We then develop a theoretical model of wave-packet evolution in time-dependent conditions. The model is applicable for a wide range of dynamical problems involving evolution in time-dependent potentials and in a changing mean-field atomic repulsion due to splitting and separation of wave-packets. 
We use the model for studying two effects that influence interferometric coherence: imperfect spatial recombination in a two-state interferometer (the so-called ``Humpty-Dumpty effect'') and phase diffusion due to number uncertainty in the two interferometer arms, which was previously studied thoroughly only for interferometric schemes where the BECs in the two arms stay trapped (for example, in a double-well potential). 
For both effects we extend the applicability of the theory to a wide range of interferometric scenarios that were not included in previous theories and provide design and optimization tools for improving the performance of matter-wave interferometers.

\end{abstract}
 
\maketitle
 
\section{Introduction}
 
Matter-wave interferometry with ultracold atoms has become a wide field of fundamental and applied research~\cite{interferometry1,interferometry2}  using many different techniques for splitting, guiding, re-combining, and probing. Coherent spatial splitting of initially trapped atoms is performed by light pulses (in Ramsey-Bord\'e~\cite{Borde1989}, Raman~\cite{ Kasevich-Chu1991,Gustavson1997,Canuel2006,WuMuller2017} or Bragg~\cite{WangCornell2005, Garcia2006,Burke2008,Hannover2012,Plotkin-Swing2018} configurations), by optical or magnetic fields that form potential barriers~\cite{Shin2004,Collins2005,Schumm2005,Jo2007} or by state-dependent  magnetic forces~\cite{Machluf2013,Margalit2015,Amit2019}.  In any of these schemes the most crucial factor is coherence, namely maintaining and retrieving a well-defined phase difference between the interferometer arms. 

Loss of interferometric coherence is commonly caused by phase fluctuations due to coupling of the atoms to the noisy environment or due to instability of elements of the device that manipulates the atoms, such as electric currents or optical fields~\cite{Stern1990,Ivanikov2018,Margalit2019}. 
However, here we wish to focus on two intrinsic effects leading to the loss of coherence that are related to entanglement between degrees of freedom of the atoms themselves:  entanglement between their spatial and internal degrees of freedom and entanglement between many atoms in the presence of atom-atom interactions. 
 
Some interferometric schemes use the internal state of the atoms for spatial splitting. 
Long before the experimental realization of interferometry with neutral atoms it was proposed to use the Stern-Gerlach effect for coherent splitting of atoms with spin $\frac12$ by applying magnetic gradients for splitting, stopping, accelerating and re-combining an atomic beam. The output signal is then the spin state of the atoms after re-combination, which depends on the spatial phase accumulated along the two interferometer arms. However, it became clear that the coherence of such an interferometer relies on a very precise spatial re-combination of the two atomic beams at the output port, such that it would be almost impossible to implement such an interferometer in practice. Theoretical investigations of this decoherence effect in a Stern-Gerlach interferometer, termed ``the Humpty-Dumpty effect", have used a simple model of a single-particle Gaussian wave-packet~\cite{Englert1988,Schwinger1988,Scully1989}. Recently such an interferometer was realized experimentally with a Bose-Einstein condensate (BEC)~\cite{Amit2019,Margalit2018} and a more elaborate model for describing its coherence is required. Note that a similar model applies for interferometer schemes that use optical pulses to split atoms. Although the spatial control in such interferometers is much more precise than in a Stern-Gerlach interferometer, imprecise spatial recombination should still be taken into account in order to optimize the performance~\cite{Roura2014}. 

Atom-atom interactions may be crucial for interferometric coherence when the atoms are trapped or guided during the sequence, when their density is relatively high. 
This effect was already noticed in the first observation of interference between two BECs~\cite{Andrews1997} and interpreted as a consequence of the predicted effect of phase diffusion due to atom-atom interactions~\cite{LewensteinYou1996,JavanainenWilkens1997,CastinDalibard1997}.  Phase uncertainty grows with the time of propagation through the interferometer arms as a result of an uncertain number difference between the two arms and hence a difference in the interaction energy. In trapped-atom interferometers phase diffusion due to interactions usually limits the coherence time to a few tens of milliseconds unless the BEC is very dilute or number squeezing takes place due to slow separation~\cite{Jo2007,Ilo-Okeke2010,Grond2010}. 
The rate of phase diffusion has been calculated for cases in which the BEC is kept in a trap or a harmonic potential along the entire interferometric sequence~\cite{LewensteinYou1996,JavanainenWilkens1997,CastinDalibard1997,Ilo-Okeke2010,Fallen2015}. 
However, it is necessary to understand the effects of number uncertainty and atom-atom interactions when the atomic wave-packets evolve non-adiabatically in confined configurations such as a matter-waveguide where coherent signals have not yet been observed despite continuous efforts. Here we show that number-dependent dynamics gives rise to new kinds of dynamic evolution of the phase uncertainty that may be promising for future applications. 
 
Theoretical studies of dynamical effects crucial for coherence such as the ones mentioned above, require accurate and efficient calculation methods for the evolution of atomic wave-packets over long times and distances that correctly reproduce the dependence of this evolution on atom-atom interactions. For relatively small atomic densities such dynamics can be approximated by Gaussian wave-packet evolution that may provide an efficient tool for examination of system performance and optimization. On the other hand, in the case of a dense BEC, such a calculation would require a numerical solution of the mean-field Gross-Pitaevskii equation (GPE)~\cite{Dalfovo1999}. 
However, for many interferometric scenarios it is impractical to solve the GPE in three dimensions for the entire interferometric sequence, especially when such a calculation needs to be iterated many times for the purpose of design and stability prediction. 
For the common case of a quadratic potential, an effective approximate solution for the GPE is provided by the ``time-dependent Thomas-Fermi approximation"~\cite{CastinDum1996,Meister2017}. 
This method starts from the static Thomas-Fermi (TF) approximation~\cite{Dalfovo1999} for a BEC in a harmonic trap, where the atom-atom interaction energy is assumed to be much larger than the kinetic energy, and continues it to the time-dependent domain. 
 
A few analytical methods were derived to bridge between the single-atom Gaussian wave-packet theory and the TF approximation for a large BEC~\cite{ Fetter1997, Fetter1998,Mateo2006,Mateo2007,Mateo2008,Nicolin2008}. These theories were mainly applied to the static state of the atoms in a trap and some were extended to specific time-dependent situations~\cite{Nicolin2008}, but none were employed for wave-packet propagation needed for atom interferometry. Another study attempted to generalize the time-dependent TF approximation for arbitrary atom-atom interaction strength~\cite{Jamison2011}, but did not go beyond the case of free expansion. 
 
Here we present a theory of atomic wave-packet evolution in a locally smooth potential that can be expanded in a quadratic form over the volume of the wave-packet. This includes evolution of the atoms under the influence of static or pulsed potential gradients or harmonic potentials for confinement or guidance, as long as the axes of the harmonic potential do not change non-adiabatically (non-rotational evolution). 
The theory reduces the Gross-Pitaevskii equation into a set of simple ordinary differential equations for the three scaling parameters as in the time-dependent TF approximation, but in addition it is valid for the entire range of atom-atom interactions and coincides with the exact Gaussian wave-packet theory in the absence of interactions (the single-atom limit). 
 
To facilitate a simple and efficient treatment of a BEC we start with a generalized TF approximation for the ground-state properties in a harmonic trap that faithfully reproduces the wave-packet size and energy over the whole range of interactions and provides a basis for presenting the dynamic properties during subsequent evolution. While some of the ingredients of this theory are already present in previous work ({\it e.g.}~\cite{Jamison2011}), we believe that only the theory presented here has now been sufficiently generalized to study a variety of interferometric schemes, and that it will be particularly useful for designing and analyzing new interferometric configurations, as demonstrated here for two major aspects of coherence. 
 
We present our wave-packet evolution method in Sec.~\ref{sec:wpprop} of this paper, and then use it as a basis for a detailed theory of decoherence due to the two intrinsic effects mentioned above. 
In Sec.~\ref{sec:spincoherence} we present the theory of coherence in a two-state spatial interferometer due to imprecise recombination at the output port. We derive explicit expressions for the expected visibility of such an interferometer with a BEC having any number of atoms and for a thermal cloud where atom-atom interactions are negligible. In Sec.~\ref{sec:phasediffusion} we present a full many-body theory of phase diffusion for trapped or propagating wave-packets. While our theory agrees with previous derivations of this effect in interferometers using spatially adiabatic splitting ({e.g. in a double-well potential) of a dense BEC, we show a different behavior of the rate of phase diffusion when the number of atoms in the BEC is relatively small. 
For propagating wave-packets we find an additional phase term that appears from the many-particle structure. Our theory may enable an analysis that can be crucial to guided matter-wave interferometry as it can help in understanding why guided interferometry has not been successful so far despite continuous efforts~\cite{ Gupta2005,Wu2007,Japha2007,Jo2007, Baker2009,Sherlock2011,Turpin2015,Navez2016,Pandey2019}. 
On the other hand, the theory may also be used to control phase diffusion to achieve very high coherence for long propagation times. 
Each section of this paper (and some sub-sections too) opens with a more detailed introduction to previous work in the field. Finally we conclude in Sec.~\ref{sec:outlook} with an outlook. 
 
\section{Wave packet evolution} 
\label{sec:wpprop}
 
This section presents the theory of wave-packet evolution necessary for describing propagation through interferometer arms either in free space (with or without additional forces, see introduction for references), or in a matter waveguide~\cite{WangCornell2005,Gupta2005,Wu2007,Japha2007,Jo2007, Baker2009,Sherlock2011,Turpin2015,Navez2016,Pandey2019} or in moving traps~\cite{Jo2007,Stevenson2015}. 
A direct analytical solution of the evolution problem is usually unavailable, while an accurate numerical solution is often impractical. One should then seek for a simplified description of wave-packet evolution that involves parametric equations. Gaussian wave-packets are very useful in light optics or matter-wave optics of non-interacting particles. 
A helpful approach for interacting matter waves is the scaling approximation,  which assumes that the shape of the wave-packet's envelope at any time remains practically the same as that of the initial wave function but with scaled coordinates and additional phase factors (see a brief review in Sec.~\ref{sec:scaling} below). 
The original scaling approximation that was applied to a BEC assumed atom-atom interactions sufficiently strong to justify the Thomas-Fermi approximation for the initial cloud in the trap, having an inverted parabolic shape with negligible kinetic energy~\cite{ CastinDum1996}. Generalizations of this method also take into account the initial kinetic energy 
in order to allow the calculation of wave-packet expansion in free space for any number of atoms~\cite{Jamison2011}. Other studies have even gone beyond the scaling approximation for calculating the expansion of a BEC in one dimension~\cite{Nicolin2008} but, to the best of our knowledge, a general theory of wave-packet evolution in the presence of time-dependent potentials relevant to interferometric scenarios and for a wide range of atom-atom interaction strengths has not been proposed. 
 
In order to further extend the effective evolution method to include evolution not only in free space but also in any potential that may be approximated locally by a quadratic form, such as a waveguide, 
or a harmonic trap changing in time, or any large-scale potential that is smooth enough over the scale of a wave-packet, we develop in this section a generalized scaling approximation for the propagation. For this method to be effective for an arbitrary number of atoms and initial trapping potentials ({\it i.e.}, a harmonic trap with any aspect ratio between the axes and strength of confinement), 
we start  with a generalized Thomas-Fermi approximation that is valid throughout this range. Although alternative methods for approximating the static or even some dynamic properties of a BEC over this range were already proposed in the past (see a brief review in Sec.~\ref{sec:GTF} below), we believe that our method is most suitable as a basis for the wave-packet evolution method in the context of atom interferometry,  which is the main subject of this work. 
 
\subsection{Generalized Thomas-Fermi approximation}
\label{sec:GTF}
 
We start with an effective mean-field theory for the static properties of the ground state of a Bose gas in a harmonic trap. 
Let us consider $N$ bosonic atoms of mass $m$ in a harmonic trapping potential $V({\bf r})=\frac12 m\sum_{j=1}^3  \omega_j^2r_j^2$, where $r_j=x,y,z$ for $j=1,2,3$, respectively, are the Cartesian coordinates parallel to the axes of the trap and $\omega_j$ are the respective trap frequencies. 
If the interaction between the atoms can be approximated by a mean-field potential then the ground state of the system is approximately a Bose-Einstein condensate (BEC): a state where all the atoms occupy the same spatial wave function $\Phi_0({\bf r})$ satisfying the stationary Gross-Pitaevskii equation (GPE) ~\cite{Dalfovo1999}
\be \left[-\frac{\hbar^2\nabla^2}{2m}+\frac{m}{2} \sum_{j=1}^3\omega_j^2r_j^2+gN|\Phi_0|^2-\mu\right]\Phi_0({\bf r})=0. \label{eq:GPE0} \ee
Here the mean-field repulsive potential (third term in the brackets) is proportional to the atom density $N|\Phi_0|^2$, where the wave function is normalized to unity, the coupling strength is $g=4\pi\hbar^2a_s/m$ with $a_s$ being the $s$-wave scattering length, and $\mu$ 
is the chemical potential. 
 
One limit of Eq.~(\ref{eq:GPE0}) is a dilute gas where the interaction term is negligible. Then  Eq.~(\ref{eq:GPE0}) reduces to a Schr\"odinger equation for a single particle, whose solutions are energy eigenstates of the single-particle Hamiltonian. The chemical potential may then be replaced by any energy eigenvalue of the harmonic trap 
$\mu\to \hbar\sum_{j=1}^3 \omega_j(n_j+\frac12)$, with $n_j$ being non-negative integers. This case of non-interacting atoms will be included in the theory of wave-packet evolution in Sec.~\ref{sec:scaling} below, but here we focus on the ground state solution. 
 In the single-particle limit the ground-state wave function $\Phi_0$ is a Gaussian. 
 
The opposite limit is the Thomas-Fermi (TF) limit, where the kinetic term (first term) is negligible relative to the potential terms. In this case the ground-state solution $\Phi_0$ has an inverted parabolic shape with a sharp edge. 
The TF approximation~\cite{Dalfovo1999} usually provides successful predictions for the state of a BEC of many atoms in a trap with a relatively low aspect ratio between its axes. Analytical theories for the ground-state properties of a BEC that extend the TF approximation beyond the strong interaction case ~\cite{Fetter1997,Mateo2006,Mateo2007,Mateo2008} or beyond the TF wave function edge~\cite{Fetter1998} have been successful in providing good approximations that agree well with the numerical solution of the GPE for different trap geometries. Variational methods were also proposed for studying the dynamics of a BEC in specific geometries~\cite{ Nicolin2008}. However, as far as we know, none of these proposals was used for calculating the dynamics of a BEC that is released from an anisotropic trap in 3D and allowed to propagate in space, as we wish to do in this work. 
 
Here we do not use a variational procedure that starts from a specific trial function such as a generalization of the trial function in Ref.~\cite{Fetter1997} to anisotropic traps. Instead, we assume that the wave function $\Phi_0$ is an implicit hybridization of an inverted parabolic wave function $\Phi_{TF}\propto \sqrt{1-\sum_j r_j^2/r_{j,{\rm max}}^2}$ that is nonzero only when the argument of the square root is positive, and a Gaussian $\Phi_G\propto \exp(-\sum_j r_j^2/4\sigma_j^2]$. 
We use the inverted parabolic form in the interaction term $gN|\Phi_0|^2$ in Eq.~(\ref{eq:GPE0}) and at the same time use the Gaussian form for estimating the kinetic term, 
namely
\begin{eqnarray}
gN |\Phi_0({\bf r})|^2 &\approx & \mu_{\rm int}\cdot {\rm max}\left\{1-\sum_j\frac{r_j^2}{r_{j,{\rm max}}^2},0\right\}, \label{eq:intterm} \\
-\frac{\hbar^2}{2m}\nabla^2\Phi_0& \approx & \sum_j \frac{\hbar^2}{4m\sigma_j^2}\left(1-\frac12\frac{r_j^2}{\sigma_j^2} \right)\Phi_0, 
\label{eq:kinterm} 
\end{eqnarray}
where $\mu_{\rm int}\approx gN|\Phi_0(0)|^2$ is the contribution of the interaction energy at the trap center to the chemical potential and ${\rm max}\{x,0\}\equiv x\Theta(x)$ ({\it i.e.} $x$ if $x>0$ and 0 otherwise). 
We emphasize that Eq.~(\ref{eq:intterm}) is not a definition of $\Phi_0$. It only means that the interaction term in the GPE is approximated by an inverted parabola whose width and  peak correspond to a normalized wave function  having the right width. This form neglects atom-atom interactions beyond the edge of the ellipsoidal volume defined by $r_{j,{\rm max}}$ but this does not exclude possible non-zero atomic density outside this ellipsoid. 
In contrast to the TF approximation we do take into account the kinetic energy term in the GPE, as represented by Eq.~(\ref{eq:kinterm}). 
 
By substituting the two terms in Eqs. ~(\ref{eq:intterm}) and~(\ref{eq:kinterm}) into Eq.~(\ref{eq:GPE0}) and equating the terms proportional to $1$ and $r_j^2$, respectively, we obtain
\begin{eqnarray}
\mu &=& \frac12 \hbar\sum_j \nu_j+\mu_{\rm int}, \label{eq:mu} \\
\omega_j^2 &=& \nu_j^2+\frac{2\mu_{\rm int}}{mr_{j,{\rm max}}^2}, \label{eq:omegaj}
\end{eqnarray}
where 
\be \nu_j=\frac{\hbar}{2m\sigma_j^2}. 
\label{eq:nuj}\ee
 
Eq.~(\ref{eq:omegaj}) is a set of 3 equations with 7 unknowns: the Gaussian widths $\sigma_j$, the ellipsoid radii $r_{\j,{\rm max}}$ and the interaction energy $\mu_{\rm int}$ at the center. In order to eliminate some of these variables we first generalize the definition of $\sigma_j$ to be meaningful for any wave function form 
\be \sigma_j^2\equiv \int d^3{\bf r}\, r_j^2|\Phi_0({\bf r})|^2.
\label{eq:sigmaj2} \ee
For a 3D inverted parabolic density as in Eq.~(\ref{eq:intterm}) the widths turn out to be $\sigma_j=r_{j,{\rm max}}/\sqrt{7}$. Different relations between $\sigma_j$ and $r_{j,{\rm max}}$ are obtained if one considers the case in which the wave function is Gaussian in some direction. For example, for a highly elongated trap where the transverse wave function is Gaussian while only the longitudinal wave function may be approximated by an inverse parabolic shape in 1D we have for the longitudinal axis $\sigma_{\parallel}=r_{\parallel,{\rm max}}/\sqrt{5}$. However, let us now take the 3D relation as a basis for the calculation and take into consideration the effect of different dimensionality as a further improvement of the approximation below. 
As a final step for eliminating the extra variables from Eq.~(\ref{eq:omegaj}) we apply the normalization condition $\int d^3{\bf r}\,|\Phi_0|^2=1$ for the inverted parabolic wave function in Eq.~(\ref{eq:intterm}) so that for 3D $\mu_{\rm int}=15 gN/8\pi \prod_j r_{j,{\rm max}}$. By including these identities in Eq.~(\ref{eq:omegaj}) we finally obtain
\be \left(\frac{\ell_j}{\sigma_j}\right)^4+\beta\frac{\ell_j^4a_s N}{ \sigma_j^2\prod_i\sigma_i}=1, 
\label{eq:selfcons} \ee
where $\ell_j\equiv \sqrt{\hbar/2m\omega_j}$ is the harmonic oscillator length along the $j$'th axis and $\beta$ is a numerical factor that we now take to be $\beta=\beta_{3D}=60/7^{5/2}=0.4628$. 
The wave function widths $\sigma_j$ that solve the coupled set of equations in Eq.~(\ref{eq:selfcons}) provide the ground-state properties such as the interaction energy $E_{\rm int}\equiv gN\langle |\Phi_0|^2\rangle$. In our inverted parabola approximation for the interaction strength  [see Eq.~(\ref{eq:omegaj})] and $r_{j,{\rm max}}=\sqrt{7}\sigma_j$ it is given for each $j$ by
\be E_{\rm int}=\frac{4}{7}\mu_{\rm int}, \quad
\mu_{\rm int}=\frac{7}{2} m(\omega_j^2-\nu_j^2)\sigma_j^2. 
\label{eq:E_int} \ee  
 
In the weak interaction limit $gN\to 0$ we recover from Eq.~(\ref{eq:selfcons}) the single-particle result $\sigma_j=\ell_j\equiv \sqrt{\hbar/2m\omega_j}$ and $\mu_{\rm int}=0$. In the TF limit the second term on the left-hand-side of Eq.~(\ref{eq:selfcons}) becomes dominant so that $\nu_j\ll \omega_j$ and we may replace 
$\sigma_j\to \sqrt{2\mu_{\rm int}/7m}/\omega_j$ and obtain the well-known expression~\cite{Dalfovo1999} $\mu_{\rm int}=[15gN\prod_i\omega_i/4m\pi]^{2/5}\cdot m/2$. 
In the intermediate range it is very easy to solve Eq.~(\ref{eq:selfcons}) numerically. 
 
The procedure described here provides fairly good agreement with the numerical solutions of the GPE (less than $\pm 5$\% error for the range of parameters shown in Figs.~\ref{fig:genTF} and~\ref{fig:genTF_1D} below). 
In practice, we use an improved approximation that takes  into account the fact that the accurate wave function does not have the inverted parabolic shape in 3D when along some of the axes the kinetic energy becomes more dominant and the shape is more like a Gaussian. Then the integration over the trap volume becomes separable along the different axes. 
For example, in the extreme limit of a highly elongated trap the kinetic energy dominates the transverse direction and the interaction takes place only along the longitudinal direction. In this 1D case the normalization condition has to be replaced by an integral over the transverse direction that gives $\beta_{1D}=6/5^{3/2}=0.5367$. For achieving high accuracy in the general case we use a simple interpolation between the 3D and 1D values 
\be \beta (\{\sigma_j\})=\beta_{1D}\left\langle\frac{\ell_j^2}{\sigma_j^2}\right\rangle_{\rm max}
+\beta_{3D}\left(1-\left\langle\frac{\ell_j^2}{\sigma_j^2}\right\rangle_{\rm max}\right), \ee
where $\langle \ell_j^2/\sigma_j^2\rangle_{\rm max}$ is an average of the ratio over the two indices $j$ where it is maximal. 
The 3D value $\beta_{3D}$ is dominant only when $\sigma_j$ is much larger than the harmonic oscillator length $\ell_j$ along at least two axes, while $\beta_{1D}$ is dominant if along two axes $\sigma_j$ is close to $\ell_j$, implying a Gaussian shape along these axes. The value of $\beta$ in Eq.~(\ref{eq:selfcons}) varies with $\sigma_j$ and is eventually determined by the solution. 
 
The interpolation procedure, which we use in the following numerical examples, leads to an accuracy of the wave function widths and energies within less than $\pm 1$\% for all the parameter ranges that were examined here. 
We consider this accuracy to be good enough for the purpose of studying the dynamical properties of atomic interferometers, which is the main purpose of this work, and therefore we will not try to further improve the accuracy or compare our approximation to previous variational or other extensions of the Thomas-Fermi approximation that claimed ``extreme accuracy"~\cite{Mateo2007} but were not followed by a theory of the dynamics in contexts similar to the present one. An alternative to our approach could always be a full solution of the GPE in order to obtain the ground-state properties and then continuing the dynamical calculation by using ground-state variables extracted from this numerical solution. The specific approach that we present in this subsection is therefore not critical for the analysis in Sec.~\ref{sec:scaling} but it provides a convenient starting point that can be easily and quickly calculated together with a clear intuitive understanding.

\begin{figure}[t!]
\includegraphics[width=\columnwidth]{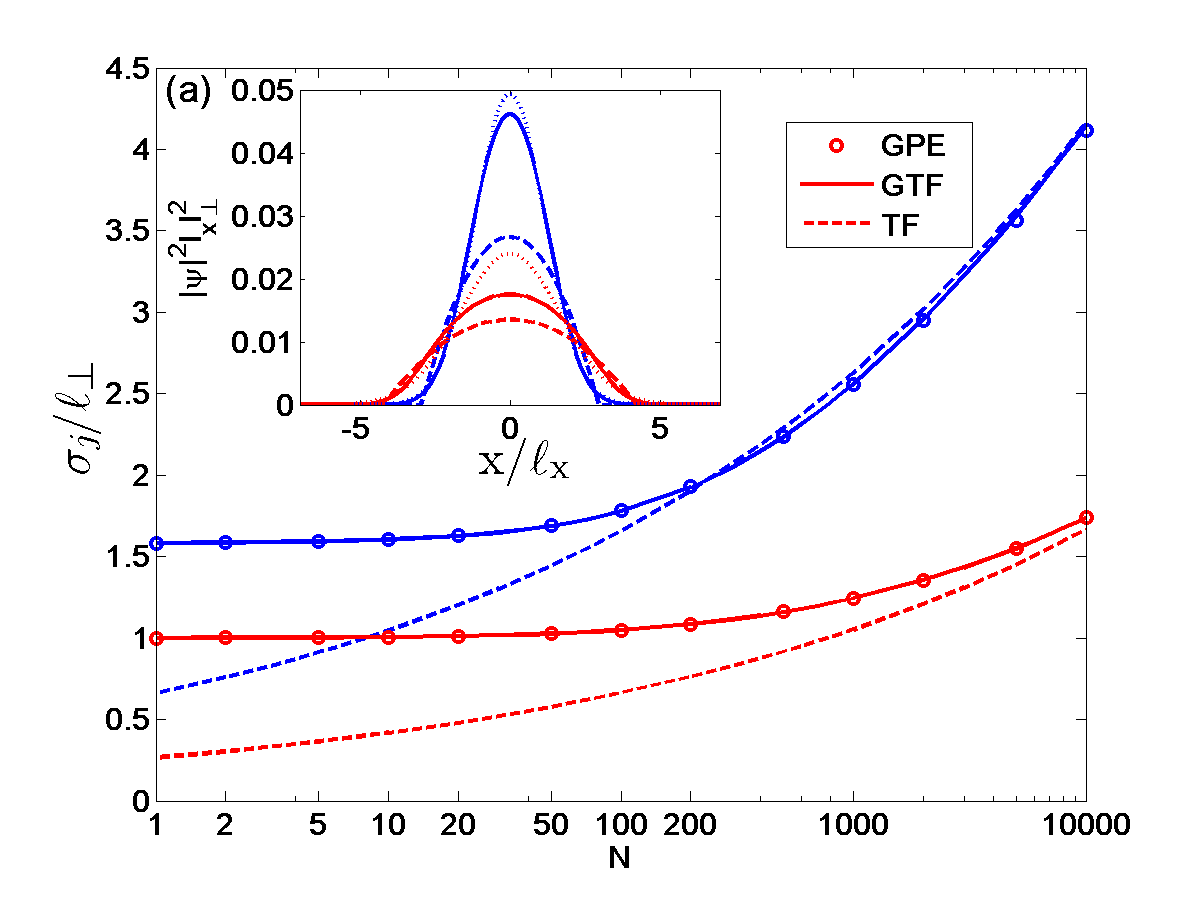} \\
\includegraphics[width=\columnwidth]{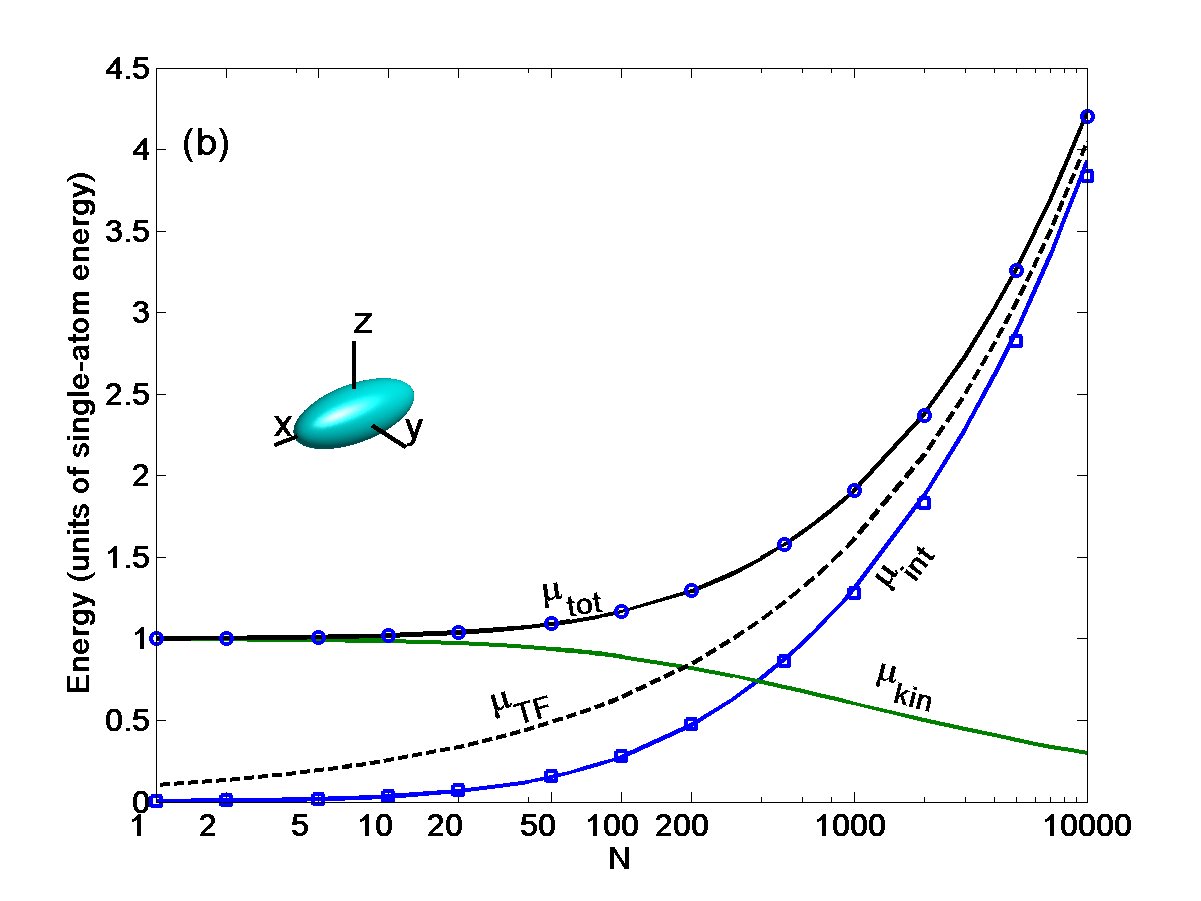}
\caption{Generalized Thomas-Fermi (GTF) approximation for the ground state of bosonic atoms in a harmonic trap. GTF (solid lines) is compared to a numerical solution of GPE (circles) and to standard TF approximation (dashed). GTF is shown to be accurate (within $\pm 1$\%) over the whole range of mean-field interaction strengths (number of atoms $N$). (a) Wave function widths $\sigma_x$ (blue) and $\sigma_{\perp}$ (red), in units of the perpendicular oscillator length $\ell_{\perp}=0.763\,\mu$m, as a function of atom number for $^{87}$Rb atoms (mass $m=1.44\cdot 10^{-25}$\,kg, $s$-wave scattering length $a_s=5.29$\,nm) in a cylindrically symmetric harmonic trap with frequencies $\omega_{\parallel}=2\pi\times 40\,$Hz and $\omega_{\perp}=2\pi\times 100\,$Hz. 
Inset: probability density profiles $|\Phi_0(x,0,0)|^2$ along the longitudinal trap axis: GPE result (solid) compared to an inverted parabolic [Eq.~(\ref{eq:intterm}), dashed] and Gaussian (dotted) profile [satisfying Eq.~(\ref{eq:kinterm})] having the same widths $\sigma_j$. For $N=100$ (blue) the GPE profile is closer to the  Gaussian, while for $N=1000$ (red) it is intermediate between the two approximate profiles. 
(b) The chemical potential due to interaction $\mu_{\rm int}$ and kinetic energy $\mu_{\rm kin}=\frac12\hbar\sum_j\nu_j$ at the trap center. The total chemical potential $\mu_{\rm tot}=\mu_{\rm int}+\mu_{\rm kin}$ according to GTF shows excellent agreement with GPE solution (circles for $\mu_{\rm tot}$ and squares for $\mu_{\rm int}=\frac{7}{4}gN\langle |\Phi_0|^2\rangle$). Inset: potential isosurface and definition of axes.}
\label{fig:genTF}
\end{figure}
 
In order to examine the validity and accuracy of our approximation in common trap geometries we compare its results to the numerical solution of the GPE [Eq.~(\ref{eq:GPE0})]. As demonstrated in Figs.~\ref{fig:genTF} and~\ref{fig:genTF_1D}, our approximation is in excellent agreement with the GPE solution over the whole range between the standard TF regime (large atom number) and the weak interaction limit (small atom number). Our approximation does not provide a prediction about the exact shape of the wave function. In the inset of Fig.~\ref{fig:genTF}(a) we present a comparison between the density profile obtained from the GPE and the  two limits of the wave-packet density profile –- a Gaussian and an inverted parabola, both having the same widths $\sigma_j$ as defined in Eq.~(\ref{eq:sigmaj2}). Although neither of the two limits of the density profile is close to the accurate profile, the GTF approximation is still successful in providing an excellent prediction for the basic properties of the wave function: size and energy, and their time evolution, as shown in the numerical examples below. 
 
\begin{figure}[t!]
\includegraphics[width=\columnwidth]{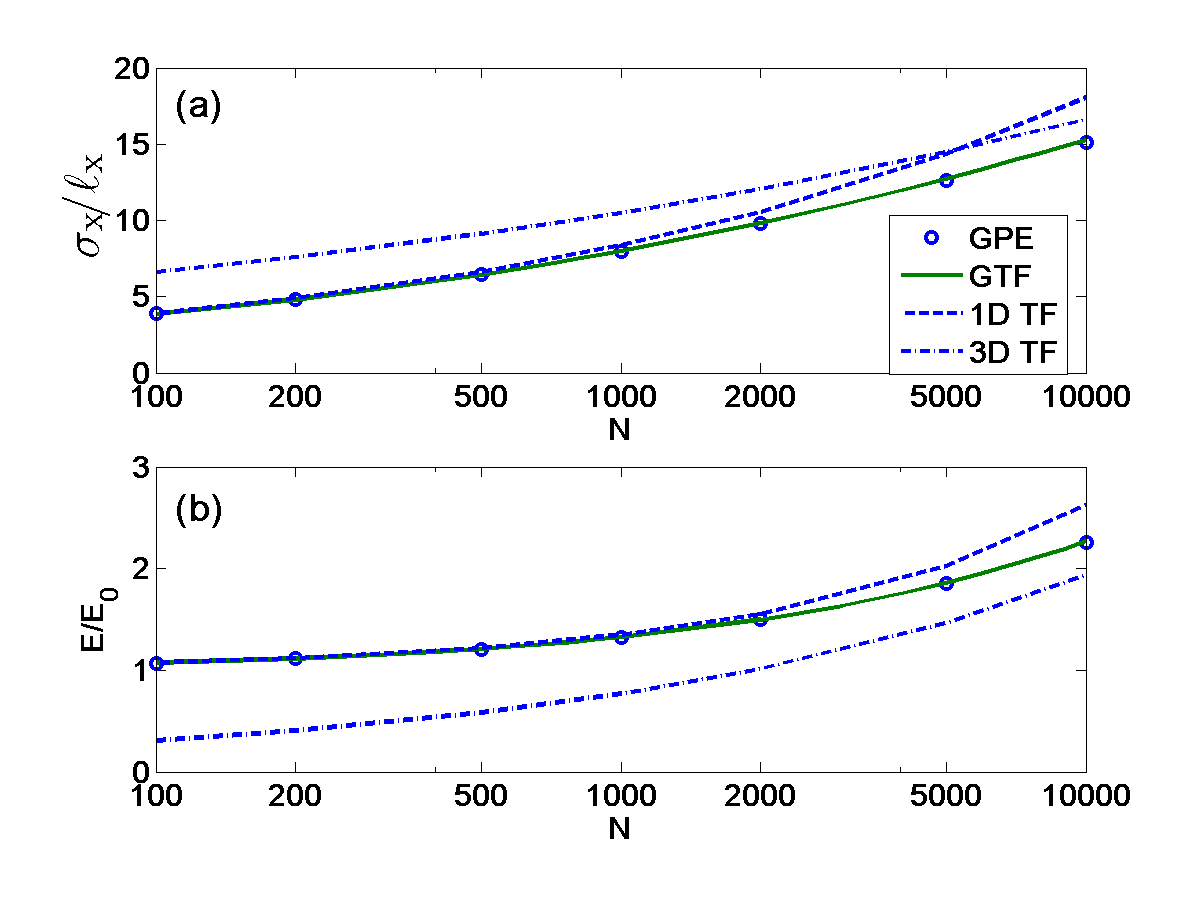} 
\caption{Generalized Thomas-Fermi (GTF) approximation for the ground state of $N$ bosonic atoms in an elongated trap ($\omega_y=\omega_z\equiv \omega_{\perp}=2\pi\times 10$\,kHz, $\omega_x=2\pi\times 40$\,Hz, other parameters as in Fig.~\ref{fig:genTF}). (a) The longitudinal cloud width (in units of the single-particle width $\ell_x=1.2\,\mu$m) and (b)  the ground-state chemical potential (in units of the single-particle energy 
$E_0=\frac12 \hbar\sum_j\omega_j$). The GTF approximation (solid lines) agrees very well with results from a numerical solution of the GPE (circles) over the entire range, while both agree with the 1D TF approximation only for low atom numbers, where the transverse wave function is the single-particle Gaussian ground state in the transverse potential. This demonstrates the validity of the GTF approximation in predicting the transition from 3D to 1D for a BEC in elongated traps. 
The condensate approximation is valid throughout the range shown at zero temperature since the $\gamma$ factor for transition into the Tonks-Girardeau regime is small.
The 3D TF approximation is valid only for $N$ beyond the range shown. }
\label{fig:genTF_1D}
\end{figure}
 
An important application of the GTF approximation is the transition from a 3D BEC to a quasi-1D bose gas in an elongated trap~\cite{Olshanii1998,Petrov2001}. In the 1D limit the large energy splitting between single-particle transverse eigenmodes of the potential allows scattering only along the longitudinal direction and hence the atomic dynamics is limited to one dimension while the wave function in the transverse direction is fixed at the lowest eigenstate of the harmonic potential. 
The physics along the longitudinal axis is then governed by an effective interaction strength $g_{1D}=g/4\pi\ell_{\perp}^2=2\hbar\omega_{\perp}a_s$~\cite{Petrov2001}. As long as the factor $\gamma=2m\omega_{\perp}/\hbar n$, where $n$ is the 1D atomic density, is small ($\gamma\ll 1$), the condensate assumption for the many-body ground state is valid (otherwise a Tonks-Girardeau gas is formed~\cite{Petrov2001,Kinoshita2004}). 
As demonstrated in Fig.~\ref{fig:genTF_1D}, the GTF approximation allows a fairly accurate prediction of the BEC properties over a broad range of parameters starting with a fully 1D BEC for low atom numbers (weak interaction) through the transition to a 3D BEC, where the interaction is strong enough to become dominant in the transverse direction.

\subsection{Evolution equations}
\label{sec:scaling}
 
Suppose that an atomic cloud is initially trapped in a harmonic potential and then at time $t>0$ the potential changes in time. For example, one may consider switching off the trapping potential along one or more axes, changing the harmonic frequencies or applying potential gradients. 
Here we treat the atoms in terms of single-atom wave-packets: a single wave-packet for a BEC, a mixture of wave-packets for a thermal cloud and two or more wave-packets for the different interferometer arms if the initial cloud is coherently split. 
We parameterize the wave-packet and derive equations of motion for the parameters, which are valid as long as the potential stays harmonic or varies smoothly in space over the volume occupied by the wave-packet. 
 
The wave function of a BEC wave-packet satisfies the time-dependent Gross-Pitaevskii  equation 
\be i\hbar\frac{\partial\psi}{\partial t}=\hat{H}_{\rm MF}(t,\psi)\psi, 
\label{eq:GPt} \ee 
where 
\be \hat{H}_{\rm MF}(t,\psi)=-\frac{ \hbar^2}{2m}\nabla^2+V({\bf r},t)+g\eta N|\psi|^2, 
\label{eq:H_GP} \ee 
and $\eta(t)N$ is the number of atoms in the wave-packet at any time. This number might change in time in different interferometric situations. For example, if a wave-packet is split into two wave-packets with an equal number of particles then after they spatially separate each one of them has $N/2$ particles and hence the strength of the mean-field potential  decreases for each wave-packet by a factor $\eta=1/2$. 
 
The following wave-packet evolution theory can also be utilized for describing the evolution of a dilute thermal cloud where atom-atom interactions can be neglected.  In this case the initial distribution in the trap may be described as a a mixture of many eigenmodes of the harmonic trap, whose evolution under the influence of the time-dependent potential can also be described with the same formalism, such that Eq.~(\ref{eq:GPt}) turns into a linear Schr\"odinger equation for each wave-packet that evolves from an initial eigenmode. As we show below, the same scaling laws apply for all these eigenmodes and therefore the treatment is quite easy and useful. 
A partially Bose-condensed atomic cloud at finite temperature where interactions are significant is beyond the scope of this work. 
 
First we consider the classical motion of the center position ${\bf R}(t)$ of the wave-packet, which evolves according to Newton's equations of motion $m\ddot{\bf R}=-\nabla V({\bf R},t)$. 
As is well known, as long as the external potential $V({\bf r},t)$ can be represented by a quadratic form, the evolution of the center-of-mass coordinates of a many-particle system can be separated from the evolution of the internal degrees of freedom of the system~\cite{Japha2002}. 
We can write the wave function as
\be \psi({\bf r},t)=e^{i[{\bf P}\cdot({\bf r}-{\bf R})+S(t)]/\hbar}\Phi({\bf r}-{\bf R},t), 
\label{eq:psirt} \ee
where ${\bf P}=m\dot{\bf R}$. By substituting this form in the evolution equation~(\ref{eq:GPt}) we obtain
the usual expression for the action $S(t)$ as an integral over the local Lagrangian
\be S=\int_0^t dt'\,\left[\frac{1}{2m}{\bf P}(t')^2-V({\bf R}(t'),t')\right], \ee
and the equaton for $\Phi({\bf r}-{\bf R})$ in the frame of reference moving with the center coordinates ${\bf R}$ becomes 
\be i\hbar\frac{\partial\Phi}{\partial t}=[H_{MF}(t,\Phi)-V({\bf R})-({\bf r}-{\bf R})\cdot\nabla V({\bf R})]\Phi, \ee
such that the 0$^{th}$ and 1$^{st}$ order terms in the expansion of $V({\bf r},t)$ around ${\bf r}={\bf R}$ are eliminated from the Hamiltonian $H_{MF}$. 
In this moving frame of reference we approximate the potential as quadratic (the next order in the Taylor expansion around ${\bf r}={\bf R}$) in a volume occupied by the wave-packet
\be V_c({\bf r}-{\bf R},t)\approx \frac12 \sum_jQ_j(t)(r_j-R_j)^2, 
\label{eq:Vc} \ee
where the quadratic potential has its axes aligned along the same axes of the initial trap, while the more general case of rotating axes is left for another work (see Ref.~\cite{Meister2017}). From here on we transform into the center-of-mass coordinate system ${\bf r}-{\bf R}\to {\bf r}$. 
 
Under the quadratic (or smoothness) condition~(\ref{eq:Vc}) we can make the scaling ansatz
\be \Phi({\bf r},t)=\frac{ \exp\left[i\left(\frac12\sum_j\alpha_jr_j^2+\varphi\right)\right]}{\sqrt{\lambda_1\lambda_2\lambda_3}}\Phi_0\left(\frac{x}{\lambda_1},\frac{y}{\lambda_2},\frac{z}{\lambda_3}\right), \label{eq:Ansatz} \ee
where $\Phi_0$ is the wave function at time $t=0$ that satisfies Eq.~(\ref{eq:GPE0}) if the initial state was a stationary state in a trap. The scaling factors $\lambda_j$, the momentum chirp $\alpha_j=\partial k_j/\partial r_j$  and the global phase $\varphi$ are time dependent and will be found below. The scaling ansatz~(\ref{eq:Ansatz}) was first used in the context of evolution of a BEC by Castin and Dum~\cite{CastinDum1996} and then by many authors (see Ref.~\cite{Meister2017} and more references therein). 
The scaling approach was originally derived for a wave function satisfying the TF approximation and was termed ``the time-dependent Thomas-Fermi method". However, one may note that the scaling ansatz is exact for the evolution of a Gaussian wave-packet in a quadratic potential in the absence of atom-atom interactions, as will be shown explicitly below. In the case of intermediate interaction strengths, as those considered in Sec.~\ref{sec:GTF} above for the stationary  problem, the scaling assumption may not be fully accurate. For example, consider a tight trap with a BEC, whose ground-state wave function satisfies the TF approximation, when the trap frequencies are lowered adiabatically. The BEC wave function is then expected to change its shape from the initial inverted parabola into a final shape closer to a Gaussian, which is the ground state in the shallow trap. 
Although the scaling assumption is not strictly satisfied, the wave-packet properties discussed in the stationary case can still be derived as time-dependent properties from the time-dependent scaling factors $\lambda_j$ of Eq.~(\ref{eq:Ansatz}), while $\Phi_0$ continues to be an implicit hybrid of the two limiting shapes.  
Here we derive the evolution equations for the scaling parameters, which are valid for the whole range of interaction strengths as in Sec.~\ref{sec:GTF}. In addition, the theory is exact for the evolution of Gaussian or Hermite-Gaussian modes that are initially eigenstates of a harmonic potential in the absence of atom-atom interaction and will therefore be valid for dilute thermal clouds that may be used for atom interferometry.

By substituting the ansatz~(\ref{eq:Ansatz}) into the left-hand side of Eq.~(\ref{eq:GPt}) and in the kinetic term we obtain
 \be \frac{i \dot{\Phi}}{\Phi}  =  -\sum_j\left(i\frac{\dot{\lambda}_j}{2\lambda_j}+ \frac{ir_j\dot{\lambda}_j}{\lambda_j}^2\frac{\partial_j\Phi_0}{\Phi_0}+\frac{\dot{\alpha}_j}{2}r_j^2\right)-\dot{\varphi}. \label{eq:dphidt} \ee
\begin{eqnarray}  
-\frac{\hbar}{2m}\frac{\nabla^2\Phi}{\Phi}  &=& -\frac{\hbar}{2m} \sum_j\left[\frac{1}{\lambda_j^2}\frac{\partial_j^2\Phi_0}{\Phi_0}+\frac{2i\alpha_jr_j}{\lambda_j}\frac{\partial_j\Phi_0}{\Phi_0}\right.  \nonumber \\
&& \left. +i\alpha_j-\alpha_j^2r_j^2\right], 
\label{eq:kin}
\end{eqnarray}
where $\partial_j$ denotes differentiation with respect to the argument $r_j/\lambda_j$ of the function $\Phi_0$. By equating the terms proportional to $r_j\partial_j\Phi_0$ in the two equations we obtain for the momentum chirp
\be \alpha_j=\frac{m}{\hbar}\frac{\dot{\lambda}_j}{\lambda_j}, \label{eq:alpha} \ee
where the relative expansion rate $\dot{\lambda}_j/\lambda_j$ along each axis may be interpreted as a  velocity chirp along the wave-packet. 
In the absence of interactions, where the initial state is a harmonic oscillator eigenstate separable into its Cartesian components, we may replace
$-(\hbar^2/2m)\partial_j^2\Phi_0/\Phi_0= \hbar\omega_j(n_j+\frac12)-\frac12 m \omega_j^2(r_j/\lambda_j)^2$, where $n_j$ is the eigenstate number. In the case of a BEC with atom-atom interactions we use the generalized TF approach and replace the first term in Eq.~(\ref{eq:kin}) by the expression in Eq.~(\ref{eq:kinterm}) with $r_j\to r_j/\lambda_j$ (which coincides with the expression for non-interacting atoms in the ground state ($n_j=0$). 
In the same spirit of the GTF, we replace the interaction term in Eq.~(\ref{eq:H_GP}) by the expression in Eqs.~(\ref{eq:intterm}) and~(\ref{eq:omegaj}). 
By collecting the terms proportional to $r_j^2$ we obtain
\[ \dot{\alpha}_j=\frac{\hbar^2}{4m\sigma_j^4\lambda_j^4}
+\frac{m}{2\hbar}\frac{\tilde{\omega}_j^2}{\lambda_1\lambda_2\lambda_3\lambda_j^2}
-\frac{\hbar}{2m}\alpha_j^2-\frac{1}{2\hbar} Q_j, \]
where
\be \tilde{\omega}_j^2=\omega_j^2-\nu_j^2 =\frac{E_{\rm in}}{2m\sigma_j^2}
\label{eq:tildew} \ee
is proportional to the interaction energy [see Eq.~(\ref{eq:E_int})], such that
$\tilde{\omega}_j\to \omega_j$ in the TF limit and $\tilde{\omega}_j\to 0$ in the interaction-free limit. 
As $\dot{\alpha}_j=(m/\hbar)\ddot{\lambda}_j/\lambda_j-(2\hbar/m)\alpha_j^2$ according to Eq.~(\ref{eq:alpha}), we obtain a differential equation for the scaling factors $\lambda_j$
\be \ddot{\lambda}_j=\frac{\nu_j^2}{\lambda_j^3}
+ \frac{\eta \tilde{\omega}_j^2}{\lambda_j \lambda_1\lambda_2\lambda_3}
-\frac{Q_j}{m} \lambda_j \label{eq:ddlambda} 
\ee
where the coefficients $\nu_j$ and $\tilde{\omega}_j$ are defined in Eqs.~(\ref{eq:nuj}) and~(\ref{eq:tildew}), respectively. 
The first term on the right-hand side of Eq.~(\ref{eq:ddlambda}) is responsible for wave-packet expansion due to position-momentum uncertainty, the second term is responsible for expansion due to the collisional repulsive force, and the third term is due to the external harmonic force (negative for $Q_j>0$ and positive for $Q_j<0$). 
In the absence of interactions $\nu_j\to \omega_j$ and $\tilde{\omega}_j\to 0$, such that the second term vanishes. In the opposite TF limit $\tilde{\omega}_j\to \omega_j$ and $\nu_j\to 0$, such that the first term in Eq.~(\ref{eq:ddlambda}) vanishes and we reproduce the result of the time-dependent TF approximation~\cite{CastinDum1996}.
 
Note that during the derivation of Eq.~(\ref{eq:ddlambda})  the only approximations that were done are the same approximations done in the derivation of the GTF approximation [Eqs.~(\ref{eq:intterm}) and~(\ref{eq:kinterm})]. It follows that Eq.~(\ref{eq:ddlambda}) is exact for the case of no interactions when the initial wave function is taken as an eigenstate of the initial harmonic potential, with any quantum number $n_j\geq 0$. 
 
An alternative form of Eq.~(\ref{eq:ddlambda}) for the wave-packet widths is obtained by taking $\sigma_j$ to be time-dependent $\sigma_j(t)=\lambda_j(t)\sigma_j(0)$, where $\sigma_j(0)$ satisfies Eq.~(\ref{eq:selfcons}) for the stationary widths of the ground state. By multiplying Eq.~(\ref{eq:ddlambda}) by $\sigma_j(0)$ and using Eq.~(\ref{eq:selfcons}) we obtain
\be \ddot{\sigma}_j=\frac{\hbar^2}{4m^2}\left(\frac{1}{\sigma_j^3}+\eta\beta \frac{a_sN}{ \sigma_j\prod_i\sigma_i}\right)-\frac{Q_j}{m} \sigma_j.
\label{eq:ddsigma} \ee
This equation does not contain explicitly any information about the initial state, but it reduces to Eq.~(\ref{eq:selfcons}) in the stationary case ($\ddot{\sigma}_j=0$, $\eta=1$)  when $Q_j/m=\omega_j^2>0$ correspond to confinement by trapping frequencies $\omega_j$.
This means that the stationary GTF theory presented in Sec.~\ref{sec:GTF} is a special case of the dynamic theory presented here. In particular, if the BEC is in its ground state in a trap then any adiabatic change of the trap frequencies leads to another stationary solution for the new ground state, which has a different relative magnitude of the kinetic and interaction energies and different corresponding wave-packet shape. 
 
Finally by collecting the remaining terms in Eqs.~(\ref{eq:dphidt}) and~(\ref{eq:kin}), which do not depend on the coordinates, together with the coordinate independent part of the interaction term in Eq.~(\ref{eq:H_GP}) we find
\be \dot{\varphi}=-\frac{1}{\hbar}\left[\frac{\eta\mu_{\rm int}}{\lambda_1\lambda_2\lambda_3}+\sum_j\frac{\hbar\nu_j}{\lambda_j^2}\left(n_j+\frac12\right)
\right], 
\label{eq:varphi} \ee
where the first term represents the mean-field effective interaction potential of a single atom under the influence of the other atoms, whose density scales with the $\lambda_j$ factors, while the second term represents the internal kinetic energy at the wave-packet center, with
$n_j$ corresponding to the mode numbers if the wave-packet is a Hermite-Gaussian function evolving from an eigenstate of a harmonic oscillator potential.

\subsection{Specific examples}
\label{sec:solutions}
 
Before applying the wave-packet evolution method to the question of interferometric coherence, we first examine its performance with some specific examples corresponding to common interferometric scenarios. Some of the following results will also be useful in the next sections, where some aspects of coherence will be discussed. 
 
\subsubsection{Free expansion}
 
One of the simplest examples of wave-packet dynamics is free expansion [$Q_j=0$ for $t>0$ in Eq.~(\ref{eq:ddlambda})]. 
If atom-atom interactions are negligible (for example, if the number of atoms is small or if the atomic cloud is dilute, as is usually the case for a thermal cloud), then $\nu_j\to \omega_j$ and the equations for the scaling factors $\lambda_j$ are simply $\ddot{\lambda}_j=\omega_j^2/\lambda_j^3$ with the solution
\be \lambda_j(t)=\sqrt{1+\omega_j^2 t^2}. 
\label{eq:free-expansion} \ee
This solution also applies to a BEC released from a cigar-shaped trap, which expands quickly along the transverse direction $y=z=r_{\perp}$, while the cloud size along the longitudinal direction stays almost fixed at short times ($\lambda_{\parallel}\equiv\lambda_x\approx 1$). In this case (for $\eta=1$) we have $(\nu_{\perp}^2/\lambda_{\perp}^3+\tilde{\omega}_{\perp}^2/\lambda_{\perp}^3\lambda_{\parallel}\approx \omega_{\perp}^2/\lambda_{\perp}^3$ with the same solution as for free expansion of non-interacting atoms, as derived previously in Ref.~\cite{CastinDum1996} in the TF approximation. 
 
Let us now consider a BEC released from a cylindrically symmetric trap and split by a quick momentum transfer into two wave-packets at time $t_0$ after release. For analyzing the evolution of the transverse wave-packet size $\sigma_{\perp}(t)=\lambda_{\perp}(t)\sigma_{\perp}(0)$ let us first denote the scaling factor just before splitting by 
$\lambda_{\perp}(t_0)\equiv \lambda_0$ and the corresponding rate of expansion by $\dot{\lambda}_0\equiv \dot{\lambda}_{\perp}(t_0)\approx \omega_{\perp}^2t_0/\lambda_0$ [see Eq.~(\ref{eq:free-expansion})]. 
After splitting (assumed to be quick such that the scaling factors do not change considerably) each wave-packet has  $\eta=1/2$ of the initial number of particles. 
The solution of Eq.~(\ref{eq:ddlambda}) for the expansion as a function of the time $t$ since the splitting is then given by
\be \lambda_{\perp}(t)=\sqrt{(\lambda_0+\dot{\lambda}_0t)^2+\lambda_0^{-2}(\eta\tilde{\omega}_{\perp}^2+\nu_{\perp}^2)t^2}. 
\label{eq:lambdaperpt} \ee
In the TF limit $\tilde{\omega}_{\perp} \to \omega_{\perp}$ and when the initial expansion time is short $t_0\to 0$, we have $\lambda_{\perp}\approx \sqrt{1+\eta\omega_{\perp}^2t^2}$. We will take this limit as a test case for phase diffusion in Section~\ref{sec:phasediffusion}.

\subsubsection{Quick splitting in a trap}
 
In order to examine the wave-packet model in scenarios where a dynamically changing atom-atom interaction plays an important role we introduce here a simple example where the number of atoms in the wave-packet 
changes drastically due to splitting. This affects the evolution of the wave-packet sizes in Eq.~(\ref{eq:ddlambda}) and its phase in Eq.~(\ref{eq:varphi}) through the fraction parameter $\eta$. This will give us an indication about the validity of the wave-packet model in interferometric situations where atom-atom interactions are important. 
 
Consider an interferometric scheme in which an initially trapped BEC is split into two parts that remain trapped, as in the Sagnac interferometer scheme proposed in Ref.~\cite{Stevenson2015}. 
Immediately after the two wave-packets separate, each of them contains only $N/2$ particles and each of them, suddenly in their own traps, no longer satisfy 
the stationary GPE. The subsequent dynamics around the center of each wave-packet is governed by Eq.~(\ref{eq:ddlambda}) with $\eta=1/2$. If the trap frequencies are not changed ($Q_j/m=\omega_j^2$), then the cloud size will first shrink due to the reduced repulsive force, initiating breathing-type oscillations in the trap around values $\lambda_j<1$ of the scaling factors.  The transverse oscillation frequency may be approximated by $2\omega_{\perp}$, as we show in Sec.~\ref{sec:waveguidesplitting}. 
In the TF limit the chemical potential, which is due only to the atom-atom interaction, 
decreases by one-half after the abrupt wave-packet separation and hence the phase at the wave-packet center starts to evolve with half the rate it had before the splitting. However, shrinking of the wave-packet and the size oscillations increase the phase change rate and it starts to oscillate. More generally, our wave-packet model predicts that the chemical potential decreases by a factor $(\frac12+b)/(1+b)$, where $b=\frac{\hbar}{2}\sum_j \nu_j/\mu_{\rm int}$ if the kinetic energy $\frac12\hbar\sum_j \nu_j$ is not negligible.

\begin{figure}[t!]
\includegraphics[width=\columnwidth]{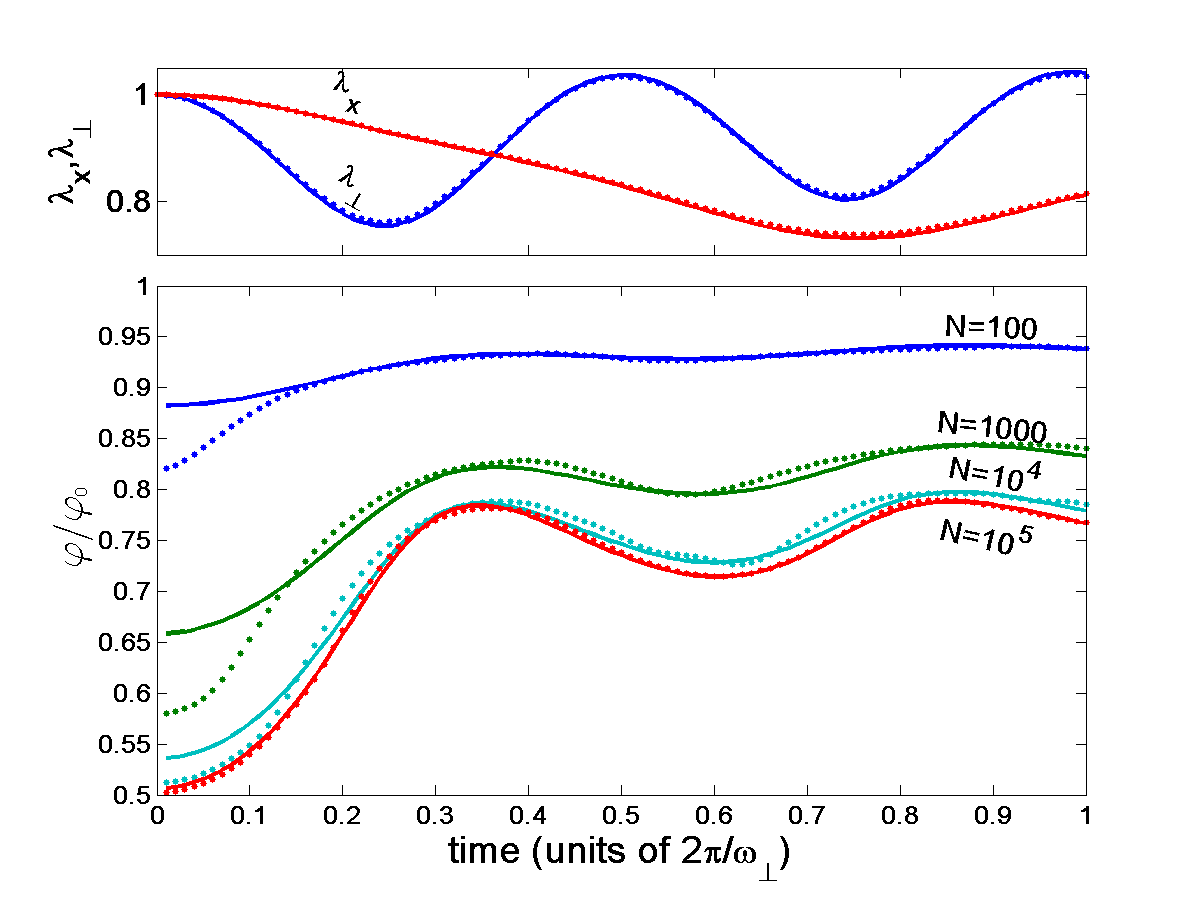}
\caption{Quick splitting into two separate traps: phase evolution after a sudden reduction of the atom number in each single trap to half the number before splitting -- comparison between wave-packet model (solid lines) and numerical GPE results (dots). Parameters before and after splitting are as in Fig.~\ref{fig:genTF}. The reduction of the repulsive mean-field potential by a factor $\eta=\frac12$ causes breathing oscillations of the BEC (top panel, $N=10^4$ atoms). The phase at the center of the wave-packet $\varphi(t)$ (bottom panel) is shown scaled by the phase of the original wave-packet $\varphi_0(t)=-\mu_{\rm tot} t/\hbar$ if it 
had not been split [$\mu_{\rm tot}$ is chemical potential given in Fig.~\ref{fig:genTF}(b)]. The wave-packet model agrees well with the numerical results except at short times (less than 1\,ms) where the accumulated phase is much less than a radian.}
\label{fig:etachange}
\end{figure}
 
In Fig.~\ref{fig:etachange} we show the wave-packet sizes and central phase $\varphi(t)$ scaled by the phase of the original wave-packet without splitting $\varphi_0(t)=-\mu_{\rm tot} t/\hbar$, over a single period of the transverse trap frequency $\omega_{\perp}$. In the case of a large number of atoms the initial phase reduces to about half that of the original phase but then, when the wave-packet shrinks and oscillates, the phase grows to a higher percentage of the unsplit wave-packet phase. The calculation based on Eqs.~(\ref{eq:ddlambda}) and~(\ref{eq:varphi}) agrees to less than about 1\% accuracy with the result of the GPE (dots) except at very short times, (less than 1\,ms) where the total phase is a small fraction of a radian. It follows that for long times relevant to interferometry the wave-packet model reproduces accurately the results of the numerical solution of the GPE. 
 
\subsubsection{Splitting in a waveguide}
\label{sec:waveguidesplitting}
 
Consider momentum splitting of a BEC in a waveguide with transverse frequency $\omega_{\perp}$ (same as the initial trapping frequency). 
Such a splitting scheme was proposed for Sagnac interferometry~\cite{Baker2009} and performed experimentally without fully eliminating the trapping potential in the longitudinal direction~\cite{WangCornell2005,Garcia2006,Burke2008}. 
The results demonstrated here will also be used in Sec.~\ref{sec:phasediffusion} for analyzing phase diffusion in waveguides. 
 
We start by studying the splitting process itself (Fig.~\ref{fig:splitting}) and then examine the long-time evolution after splitting (Fig.~\ref{fig:lambdaguide}). 
We consider a BEC prepared in a cylindrical trap with the same parameters as Fig.~\ref{fig:genTF}. The longitudinal potential is ramped down quickly to form a waveguide potential with confinement only along the transverse direction. Then quick Bragg pulses imprint a sinusoidal density grating that represents a superposition of two opposite momenta $\pm 2\hbar k$, where $k$ is the wave-vector of the Bragg laser. Here we ignore the specific atom-light interaction scheme, which can be found in the literature~\cite{WangCornell2005,Giese2013}, and take it as a black box generating a transformation $\psi_0({\bf r})\to \psi_0({\bf r})[e^{2ikx}+e^{-2ikx}]/\sqrt{2}$. This causes a separation of the two wave-packets, which propagate with velocities $\pm v=\pm 2\hbar k/m$. The atomic density within the interference fringe pattern that is formed in the overlap region between the two wave-packets before they separate is responsible for an enhanced collisional repulsion force along the transverse directions, which is larger than what would be expected if the density was uniform along $\hat{x}$ (for example, if the two wave-packets have two different spin states and do not interfere). In this overlap region the density is $|\psi(x)|^2\propto \cos^2(2kx)$ and hence the repulsive force is enhanced by a factor $\int_0^{\pi/k}dx\,\cos^4(2kx)/\int_0^{\pi/k}dx\,\cos^2(2kx)=3/2$. During the separation process the region of overlap with enhanced repulsion becomes smaller by a factor $e^{-(vt/\tilde{\sigma}_x)^2}$, where $\tilde{\sigma}_x$ is the effective width of each of the wave-packets, and the averaged repulsive interaction decreases exponentially to half of the original wave-packet with $N$ particles when the separation is complete. 
 
\begin{figure}[t!]
\includegraphics[width=\columnwidth]{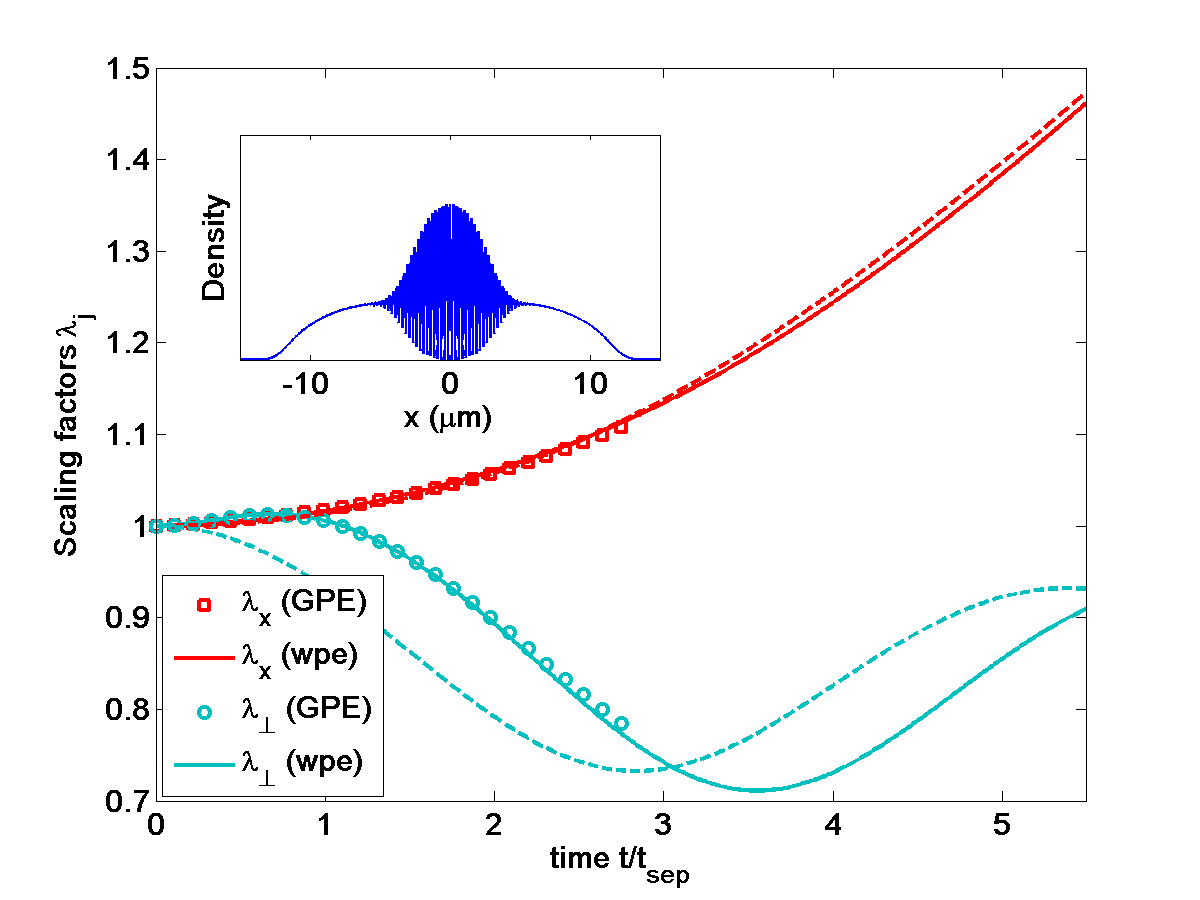}
\caption{Cloud sizes (scaling factors $\lambda_j$) after momentum splitting of a BEC in a waveguide: comparison between GPE and the wave-packet evolution (wpe) method of this work. A BEC of $N=10^4$ atoms is prepared in a cylindrical trap with $(\omega_x,\omega_{\perp})=2\pi\times (40,100)$\,Hz (same parameters as in Fig.~\ref{fig:genTF}). 
The longitudinal frequency is then switched off to form a waveguide potential along $x$ and Bragg pulses create a superposition of two opposite momenta $\pm mv=\pm 2\hbar k$, where $k=2\pi\times 1\,\mu$m$^{-1}$, such that $\psi_0({\bf r})\to \psi_0({\bf r})(e^{2ikx}+e^{-2ikx})/\sqrt{2}$. Consequently the two wave-packets start to separate while maintaining a fringe pattern in the overlap region (the inset shows the density at $t=0.5$\,ms). The wave-packets have a longitudinal extent of $\pm x_{\rm max}\approx \pm 8.4\,\mu$m, corresponding to full separation at $t_{\rm sep}=x_{\rm max}/v\approx 0.91$\,ms. The effective atom-atom repulsion in Eq.~(\ref{eq:ddlambda}) is modeled by two effective fraction factors $\eta_j(t)=\frac12+\delta\eta_je^{-(vt/\tilde{\sigma}_x)^2}$, where $\delta\eta_{\perp}=1$, $\tilde{\sigma}_x(t)=1.09\sigma_x(0)\lambda_x(t)$ is the effective wave-packet longitudinal size that determines the overlap (1.09 corrects for the non-Gaussian shape), and 
$\delta\eta_x=1/4$  provides fair agreement for the longitudinal size calculated with the GPE (red solid line).  See
text for more details. We continue the model curves beyond $t=2t_{\rm sep}$, where GPE results are not calculated, to compare them with the dashed curves, calculated with a simple model where $\eta=\frac12$ from $t=0$. This demonstrates that the long time evolution is not sensitive to the details of the splitting, which determine mainly the phase of the transverse size oscillations but not their mean or amplitude.}
\label{fig:splitting}
\end{figure}
 
Although the wave-packet model of this work is not strictly valid when the two wave-packets in the two interferometer arms partially overlap, we demonstrate here that an effective model for this situation is still useful, even in the case where the density in the overlap region  is corrugated due to interference. 
We model the repulsion by different time-dependent atom fraction factors $\eta\to \eta_{\perp}(t),\, \eta_x(t)$ in  Eq.~(\ref{eq:ddlambda}) for the transverse and longitudinal directions, respectively, such that $\eta_j(t)=\frac12+\delta\eta_j e^{-(vt/\tilde{\sigma}_x)^2}$. Here the effective wave-packet width $\tilde{\sigma}_x$ in the direction of splitting, representing the rate of drop of the overlap integral between the two wave-packets when the displacement $vt$ grows,  is a bit different from the width $\sigma_x$ defined in Eq.~(\ref{eq:sigmaj2}), due to the non-Gaussian wave-packet shape ($\tilde{\sigma}_x/\sigma_x\sim 1.09$ for an inverse parabola.  The transverse fraction factor is $\eta_{\perp}=3/2$ ($\delta\eta_{\perp}=1$), as explained above. 
The repulsion dynamics along the $\hat{x}$ direction in the overlap region is more intricate and we model it by $\delta\eta_x=1/4$, which yields fair agreement with the numerical GPE results in Fig.~\ref{fig:splitting}. The transverse size of the BEC first grows due to the enhanced repulsion in this direction but then shrinks due to the reduced repulsion relative to the harmonic force. 
The dashed curves in Fig.~\ref{fig:splitting} represent the results of a simplified model where $\eta=\frac12$ during the whole evolution starting just after the Bragg pulses. 
This model yields the same oscillation of the transverse wave-packet size but with a different phase, while the longitudinal size is almost the same as the one calculated with the detailed model. This demonstrates that the long-time evolution is not sensitive to the details of the atom-atom interaction before full separation, allowing for modeling long interferometric sequences without specifically caring about the evolution during the period when the wave-packet model is not strictly accurate.

\begin{figure}[t!]
\includegraphics[width=\columnwidth]{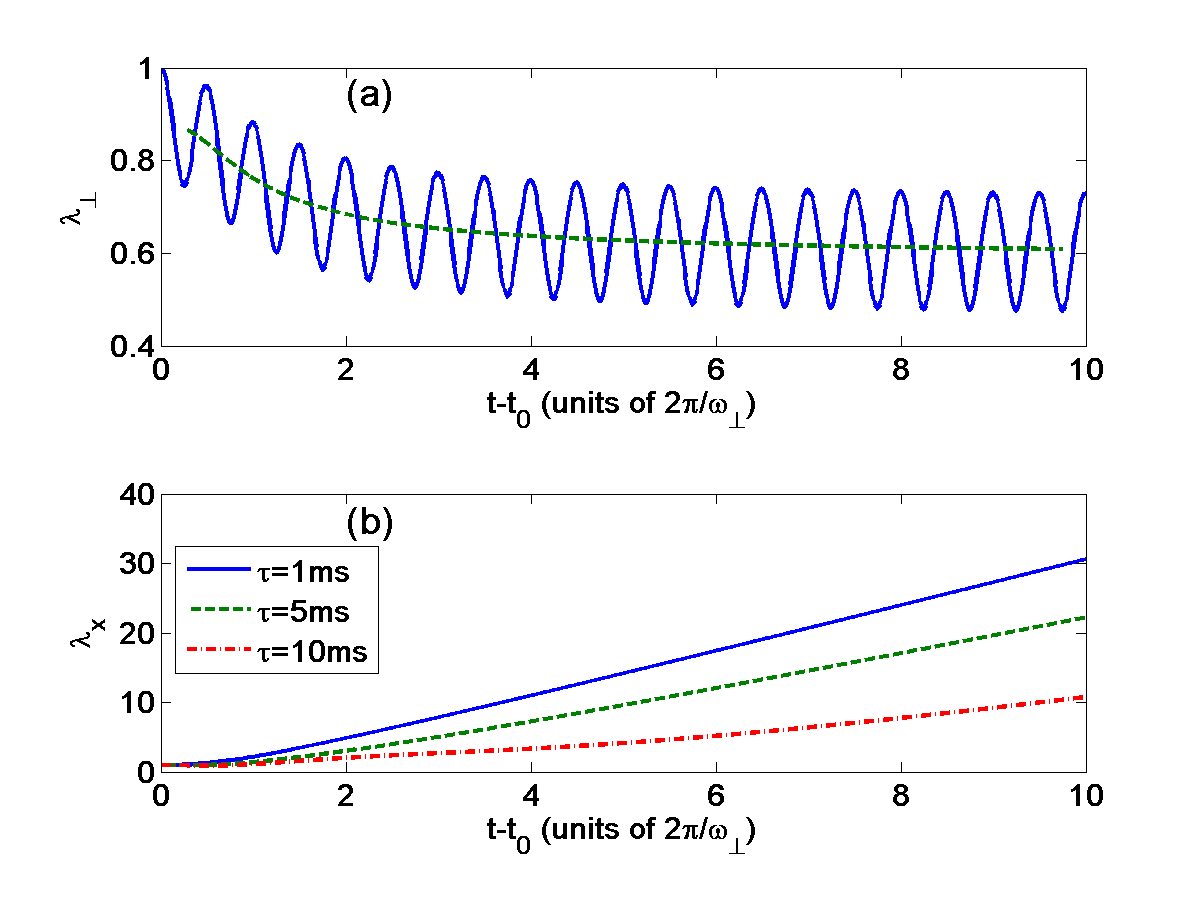} 
\caption{Long-time cloud size evolution during release, splitting, and expansion into a matter-waveguide. The procedure and parameters are as in Fig.~\ref{fig:splitting}, except that the longitudinal trapping potential is ramped down gradually as $\omega_x(t)=\omega_x(0)e^{-t/\tau }$ and the splitting is performed at $t_0=2\tau$.  
The calculation uses a simple wave-packet evolution model where the fraction factor  decreases abruptly from $\eta=1$ to $\eta=\frac12$ at $t=t_0$. 
 (a) The transverse scaling factor $\lambda_{\perp}(t)=\sigma_{\perp}(t)/\sigma_{\perp}(0)$ (for $\tau=1$\,ms) oscillates due to the sudden decrease of the atom-atom repulsive force after splitting. The oscillation with frequency $2\omega_{\perp}$ is around $\bar{\lambda}_{\perp}$ given in Appendix~\ref{app:BECwaveguide} [Eq.~(\ref{eq:barlambda})]  (dashed line).  
(b) Longitudinal scaling factor $\lambda_x(t)=\sigma_x(t)/\sigma_x(0)$ for different release times. The asymptotic expansion rate is inversely proportional to the initial cloud size and therefore it decreases when the cloud is allowed to expand slowly within a time $\tau $ before full release. 
For abrupt release ($\omega_x\tau\ll 1$) the asymptotic expansion rate is expected to be $\dot{\lambda}_x\sim 0.42$\,ms$^{-1}$; it is found to be 80\% of this value for $\tau=1$\,ms and less for longer release times, approaching  $\dot{\lambda}_x\sim 1/\tau$ for $\omega_x\tau \gg 1$. 
See Sec.~\ref{sec:phasediffusion} and Fig.~\ref{fig:guiderelease} for implications of this evolution.}
\label{fig:lambdaguide}
\end{figure}
 
The subsequent evolution after full separation is characterized by expansion in the longitudinal direction along the waveguide and oscillations of the cloud size in the transverse direction, as demonstrated in Fig.~\ref{fig:lambdaguide}. We compare the numerical solution of Eq.~(\ref{eq:ddlambda})  with analytical expressions derived in Appendix~\ref{app:BECwaveguide}. For the long-time evolution of the cloud size in Fig.~\ref{fig:lambdaguide} we use the same parameters as in Fig.~\ref{fig:splitting}, except that here we ignore the details of the splitting process at short times and set $\eta=1/2$ for $t>0$. These details are found to have a small effect on the evolution at long times $t\gg \tau_{\rm sep}\equiv m\sigma_x/2\hbar k$, where $\tau_{\rm sep}$ is the time it takes for the wave-packets to separate. 
We also examine the effect of trap release time $\tau$ on the expansion dynamics. The longitudinal frequency is ramped down as $\omega_x(t)=\omega_x(0)e^{-t/\tau}$ and the splitting is performed at $t=t_0=2\tau $. 
The asymptotic longitudinal expansion rate is proportional to the initial longitudinal trap frequency if $\tau<\omega_x^{-1}$ and to the inverse of the release time $\tau^{-1}$ if the release time is longer than the initial trap period. 
These results have significant implications on phase diffusion determining the coherence of the interferometry, as we discuss in Sec.~\ref{sec:phasediffusion}. 
 
Before concluding this section we note that calculations of BEC splitting in a waveguide have already been done in the past~\cite{Ilo-Okeke2010,Fallen2015}. These calculations, which involved a comparison between a GPE calculation and two limits of atom-atom interaction, namely the TF approximation and perturbation theory, aimed at understanding an experiment where the longitudinal potential was not turned off so that the atomic clouds moved in a harmonic potential~\cite{Garcia2006,Burke2008}. Our method is suitable for efficient calculations of dynamics in an interferometer in a broad range of possible applications such as a Sagnac interferometer in a ring waveguide in all interaction regimes including a non-interacting thermal cloud or a BEC in either the TF limit or for weak interactions.

\section{Coherence of a spatial two-state interferometer}
\label{sec:spincoherence}
 
In many interferometer schemes the atoms travel along the two interferometer arms in orthogonal internal states.  For example, the beam splitters in the Ramsey-Bord\'e~\cite{Borde1989} and Kasevich-Chu~\cite{Kasevich-Chu1991} configurations use a simultaneous transfer of momentum from a laser photon together with an internal state transition induced by the same photon. 
An archetype of such an interferometric scheme,  which was envisioned in the early days of quantum mechanics, is based on the Stern-Gerlach effect, where a magnetic field gradient turns a superposition of two spin states into a superposition of spatial paths~\cite{Machluf2013,Margalit2018}. In contrast to interferometers based on the double-slit scheme, where spatial interference fringes due to a recombination of two indistinguishable paths are observed, two-state interferometers measure the internal state of the atoms when their spatial state is recombined in position and momentum. 
The challenge of a two-state interferometer is to split an atomic wave-packet into two paths with a macroscopic separation and then bring the two wave-packets back to the same position and momentum with microscopic precision so that the two arms cannot be distinguished by their spatial state. In the framework of the Stern-Gerlach interferometer (SGI), 
erasing the entanglement between the spatial degrees of freedom and the internal degrees of freedom was considered to be a very difficult task that requires accurate manipulation of magnetic fields that can be hardly achieved by macroscopic experimental devices~\cite{Englert1988,Schwinger1988,Scully1989}, so it was termed the ``Humpty-Dumpty effect''. This challenge is successfully overcome by  matter-wave interferometers based on Raman or Bragg momentum transfer from laser photons having a very high momentum precision that does not depend on the laser intensity or duration of the pulses. Yet, imprecision effects are still important for the performance of these interferometers at large separation distances, as well as for interferometers based on splitting and guiding by continuous forces such as Stern-Gerlach interferometers~\cite{Machluf2013,Margalit2018} or interferometers using guiding potentials~\cite{Stevenson2015}. 
While the original theoretical work~\cite{Englert1988,Schwinger1988,Scully1989} that investigated the required precision of differential forces in two-state interferometers used simplifying assumptions about the symmetry of the interferometer, more recent work has investigated the effects more thoroughly in the context of light-pulse interferometers~\cite{Roura2014}.
 
Here we develop a more general theory based on our method of wave-packet evolution presented in Sec.~\ref{sec:wpprop}. This theory will be relevant to non-interacting thermal atomic clouds as well as BEC clouds with any strength of atom-atom interaction, provided that the interactions during splitting and recombination can be absorbed into parameters of the theory as in the example given in Fig.~\ref{fig:splitting}. 
The theory will enable practical calculation of interferometric performance not only in interferometers employing two internal atomic levels but also in interferometers employing momentum transitions induced by Bragg pulses, as we show below.   
 
\subsection{General result for pure state input}
In a two-state interferometer with a pure state input the atomic wave function after splitting and before recombination is a superposition of two wave-packets
\be |\psi(t)\rangle = \frac{1}{\sqrt{2}}\left[\psi_1({\bf r},t) |1\rangle+\psi_2({\bf r},t) |2\rangle\right], 
\label{eq:intstate} \ee
where $\psi_a({\bf r})$ ($a=1,2$) has the form of Eq.~(\ref{eq:psirt}) with corresponding central positions ${\bf R}_a$, central momenta ${\bf P}_a$ and central phase $S_a/\hbar$ accumulated along the interferometer arms, while the wave functions $\Phi_a({\bf r}-{\bf R}_a,t)$ have the form of Eq.~(\ref{eq:Ansatz}) in the scaling approximation. 
The internal atomic states $|1\rangle$ and $|2\rangle$ are two spin states (Zeeman states) in interferometers is based on magnetic forces or hyperfine states in Ramsey-Bord\'e or Kasevich-Chu interferometers. 
The following discussion based on Eq.~(\ref{eq:intstate}) also applies to interferometers based on Bragg transitions, where the wave-packets propagating through the two arms have two orthogonal momentum states $|1\rangle\to |2n_1\hbar k\rangle$, $|2\rangle\to |2n_2\hbar k\rangle$ with the same internal atomic state but different results of absorption-emission of photon pairs. 
The beam splitters at the input and output port of all these interferometric schemes involves an internal state rotation $|1\rangle\to (|1\rangle+|2\rangle)/\sqrt{2}$, $|2\rangle\to (|1\rangle-|2\rangle)/\sqrt{2}$ and the interferometric signal is the population in one of the internal states compared to the other. This procedure of rotation and measurement is equivalent to projecting the interferometer state in Eq.~(\ref{eq:intstate}) at the output time $t_f$ into one of the superposition states $(|1\rangle\pm |2\rangle)/\sqrt{2}$ and probing the resulting population  (or probability)
\be {\cal P}_{\pm}=\frac{1}{2}\left[1\pm V\cos(\delta\varphi)\right], \ee
where the visibility $V$ and phase $\delta\varphi$ are, respectively, the absolute value and phase of the overlap integral 
\be Ve^{-i\delta\varphi}\equiv \langle\psi (t_f)|\hat{\rho}_{12}|\psi(t_f)\rangle= \int d^3{\bf r}\,\psi_1^*({\bf r},t_f)\psi_2({\bf r},t_f). 
\label{eq:overlap} \ee
where  $\hat{\rho}_{12}\equiv |1\rangle\langle 2|$ is a projection operator. 
 
The visibility drops  if the wave-packets at the output port are displaced in position (by $\delta{\bf R}={\bf R}_1-{\bf R}_2$), or in momentum (by $\delta{\bf P}={\bf P}_1-{\bf P}_2$), or if they differ in size (scaling parameters $\lambda_j$) or in momentum chirp $\alpha_j$. 
Here we focus on the effect of imperfections that lead mainly to relative position or momentum displacements between the two arms, while the sizes and expansion rates of the two wave-packets are assumed to be equal. We generalize the treatment of the original work about the ``Humpty-Dumpty effect"~\cite{Englert1988,Schwinger1988,Scully1989} to beyond non-expanding Gaussian wave-packets and discuss the effect of expansion and atom-atom interactions. 
A similar treatment was presented in a more recent work~\cite{Roura2014} but here the results are further generalized to the case of a BEC with arbitrary atom-atom interactions. 
 
Before deriving specific expressions for the visibility and phase of the interferometric signal, it is important to note an important general property of the overlap integral in Eq.~(\ref{eq:overlap}): it is invariant under unitary operations that are independent of the internal atomic state.  This means that the the overlap integral does not change in time during free propagation or under the influence of state-independent forces. 
Consider a unitary time-evolution operator $\hat{U}$, such that $\psi_a({\bf r},t+\tau)=\hat{U}(\tau)\psi_a({\bf r},t)$. Since $\hat{U}$ is independent of the internal state, $[\hat{U},\hat{\rho}_{12}]=0$, we have $\hat{U}^{\dag}\hat{\rho}_{12}\hat{U}=\hat{U}^{\dag}\hat{U}\hat{\rho}_{12}=\hat{\rho}_{12}$, so that
\begin{eqnarray}  \langle \psi(t+\tau)\hat{\rho}_{12}\psi(t+\tau)\rangle &=& \langle \psi(t)\hat{U}^{\dag}\hat{\rho}_{12}\hat{U}\psi(t)\rangle \nonumber \\
&=& \langle\psi(t)\hat{\rho}_{12}\psi(t)\rangle. 
\label{eq:Vconserve} \end{eqnarray}
This result will be used in the following derivation. Practically it is relevant to Stern-Gerlach interferometers, where the final projection that measures the spin state does not involve momentum transfer, and it implies that the visibility and measured phase are independent of the timing of this projection, as long as spin-dependent forces do not exist at this time. 
 
To evaluate the overlap integral in Eq.~(\ref{eq:overlap}) with the wave functions having the form of Eqs.~(\ref{eq:psirt}) and ~(\ref{eq:Ansatz}) with the same scaling factors $\lambda_j$ and phase curvatures $\alpha_j$ for the $\psi_1$ and $\psi_2$ , we define ${\bf R}_{1,2}={\bf R}\pm \delta {\bf R}/2$ and ${\bf P}_{1,2}={\bf P}\pm \delta{\bf P}/2$ and transform the integration coordinates into the scaled coordinates $x_j\equiv (r_j-R_j)/\lambda_j$ centered at the center of mass ${\bf R}=\frac12({\bf R}_1+{\bf R}_2)$ of the two wave-packets. 
We then obtain
\be \langle \psi|\hat{\rho}_{12}|\psi\rangle
=  e^{-i\delta\varphi}\int d^3{\bf x}\, e^{-i\delta\bar{\bf P}\cdot{\bf x}/\hbar}\Phi_0({\bf x}-\frac12\delta{\bf X}) \Phi_0 ({\bf x}+\frac12\delta{\bf X}), 
\label{eq:overlap1} \ee
where the scaled center-to-center displacement and the scaled effective momentum difference are 
\be \delta X_j=\delta R_j/\lambda_j \quad \delta\bar{P}_j=\lambda_j
\delta P_j-m\dot{\lambda}_j\delta R_j
\label{eq:bardP}, \ee
and the interferometric phase is 
\be \delta\varphi=\phi_1-\phi_2-{\bf P}\cdot \delta{\bf R}/\hbar, 
\label{eq:deltavarphi} \ee
with 
\be \phi_a=\frac{1}{\hbar}S_a+\varphi_a \ee
for $a=1,2$. The phase accumulated along each interferometer arm includes the action along the trajectory  and the internal wave-packet phase [Eq.~(\ref{eq:varphi})] due to kinetic and interaction energy relative to the wave-packet center. The last term in Eq.~(\ref{eq:deltavarphi}) is often called ``the separation phase" due to the separation between the two end-points of the trajectories~cite{Bongs2006}. Together with this term the interferometric phase is invariant under free evolution, as required by Eq.~(\ref{eq:Vconserve}), aince $\delta S(t+\tau)-\delta S(t)=(P_1^2-P_2^2)\tau/2m={\bf P}\cdot{\bf \delta v}\tau$ is exactly opposite to the change of $-{\bf P}\cdot{\bf \delta R}$ over the time $\tau$.

The decomposition of the overlap integral in Eq.~(\ref{eq:overlap1}) into a phase factor $e^{-i\delta\varphi}$ and a real visibility $V$ is justified when the initial wave function symmetric or antisymmetric under inversion, $\Phi_0(-{\bf x})=\pm\pm\Phi_0({\bf x})$, such that the integral  must be real. 
Note that Eq.~(\ref{eq:overlap1}) has the same form as Eq.~(12) of Ref.~\cite{Roura2014}, except that here we give the explicit time-dependence of the effective momentum and position displacement in terms of the initial wave-packet envelope and the scaling factors $\lambda_j$, which are relevant to wave-packet dynamics with or without non-linear atom-atom interactions, as derived in Sec.~\ref{sec:wpprop} above. 
 
\subsection{Gaussian approximation}
If the two wave-packets at the output port ($t=t_f$) are displaced only by momentum ($\delta{\bf R}=0$) then the visibility is the Fourier transform of the initial probability density $|\rho({\bf x})=|\Phi_0({\bf x})|^2$, such that if it has spatial widths $\sigma_j$ then the visibility scales like $V\approx \exp\left(-\frac12\sum_j (\lambda_j\sigma_j\delta P_j/\hbar)^2\right)$. In the opposite case where the two wave-packets are only displaced in position and if they have a minimal size at the time of recombination ($\delta{\bf P}=0$, $\dot{\lambda}_j=0$), then the visibility drops with displacement as $V\approx \exp\left(-\frac12\sum_j (\delta R_j/2\sigma_j\lambda_j)^2\right)$. In the more general case of both position and momentum displacements we apply the principle of invariance of the overlap integral  [Eq.~(\ref{eq:Vconserve})].  The overlap integral can be written as $\int d^3{\bf x}\, \Phi_+^*({\bf x})\Phi_+({\bf x})$, where $\Phi_{\pm}({\bf x})\equiv e^{\pm i\delta\bar{\bf P}\cdot{\bf x}/2\hbar}\Phi_0({\bf x}\pm\delta{\bf X}/2)$. We apply on both wave functions $\Phi_{\pm}$ the unitary operator 
$\hat{U}({\bf \theta})=\prod_j \exp\{-i\theta_j[(\sigma_j\hat{p}_j/\hbar)^2+(\hat{x}_j/2\sigma_j)^2]\}$, where $\hat{p}_j$ are the momentum operators conjugate to the position operators $\hat{x}_j$ and $\sigma_j$ are the widths of the initial wave function $\Phi_0$.  
The operation of $\hat{U}$ creates a phase space rotation of the center coordinates as in a system of harmonic oscillators with frequencies $\nu_j=\hbar/2m\sigma_j^2$  [see Eq.~(\ref{eq:nuj})]. 
The operation $\hat{U}$ rotates the wave-packet center phase space coordinates
\[ \delta X_j\to \cos\theta_j\delta X_j-\frac{1}{m\nu_j}\sin\theta_j\delta \bar{P}_j, \], 
\[ \delta\bar{P}_j\to m\nu_j\sin\theta_j\delta X_j+\cos\theta_j\delta\bar{P}_j, \]
 while the shape of $\Phi_0$ in the frame moving with the center coordinates is conserved by the rotation if $\Phi_0$ is a Gaussian or Gaussian-Hermite eigenstate of the harmonic oscillator Hamiltonian. If $\Phi_0$ is not an eigenstate but rather a stationary solution of the GPE  for interacting atoms, then the operation $\hat{U}$ may change the wave function shape but conserve the widths $\sigma_j$. In the scaling approximation this corresponds to stationary scaling factors satisfying Eq.~(\ref{eq:ddlambda}) with $\eta=0$ and $Q_j/m=\nu_j^2$.
We then choose $\theta_j={\rm atan}[m\nu_j\delta X_j/\delta\bar{P}_j]$ such that $\delta X_j\to 0$, while $\delta\bar{P}_j\to \sqrt{\delta\bar{P}_j^2+(\hbar\delta X_j/2\sigma_j^2)^2}$. 
 
By applying this transformation to Eq.~(\ref{eq:overlap1}) the visibility reduces to 
\be V\simeq \int d^3{\bf x}\,e^{-i\sum_{j=1}^3 x_j\sqrt{(\delta\bar{P}_j/\hbar)^2+(\delta X_j/2\sigma_j^2)^2}}|\Phi_0({\bf x})|^2, 
\label{eq:overlapU} \ee
which is exact if $\Phi_0$ is a Gaussian. For such a Gaussian wave function or as a Gaussian approximation for other wave functions we obtain
\be V=e^{-\frac12\sum_{j=1}^3\left[(\sigma_j\delta\bar{P}/\hbar)^2+(\delta X_j/2\sigma_j)^2\right]} 
\label{eq:overlap_pure} \ee
with $\delta X_j$ and $\delta\bar{P}_j$ defined above in Eq.~(\ref{eq:bardP}). 
For a 3D inverted parabolic wave function (TF limit) Eq.~(\ref{eq:overlap1}) yields for a pure position displacement along a given axis $V(\delta X_j,\delta\bar{\bf P}=0)\approx e^{-(\delta X_j/2\zeta_x\sigma_j)^2/2}$ with $\zeta_x=0.8267$, while for a pure momentum displacement we obtain $V(\delta {\bf X}=0,\delta P_j)\approx e^{-(\zeta_p\sigma_j\delta P_j/\hbar)^2/2}$ with $\zeta_p=1.093$ (obtained by numerical integration). For an accurate estimation with arbitrary displacements one must perform a direct numerical integration of Eq.~(\ref{eq:overlap1}) . 
 
Let us note that the law of conservation of visibility [Eq.~(\ref{eq:Vconserve})] can be verified explicitly by checking that $\frac{d}{dt}[\delta\bar{P}_j^2+(m\nu_j\delta X_j)^2]=0$ if the evolution of $\delta P_j$ and $\delta R_j$ satisfies Newton's equations of motion and the evolution of  $\lambda_j$ satisfies Eq.~(\ref{eq:ddlambda}) with no interactions. 
However, the conservation of overlap is not explicitly satisfied in the presence of collisional interactions ($\eta\neq 0$), as our approximation in Sec.~\ref{sec:wpprop} does not take into account the interaction between different wave-packets. This means that the effective potential according to this model is different for the two wave-packets, as each one of them experiences a mean field repulsion only due to its own density. 
 
\subsection{Mixed state input and application} 
 Now let us consider an initial mixed state represented by a density matrix
\be \rho_0({\bf r},{\bf r}')=\sum_n W_n\Phi_n({\bf r})\Phi_n^*({\bf r}'), \ee
where $W_n$ are weights ($\sum_n W_n=1$) and $\Phi_n$ are a set of orthogonal eigenstates of the initial trapping potential. 
The interferometric process splits each wave function into a superposition $\Phi_n\to \frac{1}{\sqrt{2}}(\psi_n^{(1)}+\psi_n^{(2)})$, where $\psi_n^{(a)}({\bf r},t=0)=\Phi_n({\bf r})$ (for $a=1,2$) but then change in time in an internal-state dependent potential until the arms are  recombined.  The output signal is then
\begin{eqnarray} {\cal P}_{\pm} &=& \sum_n W_n\sum_{a,b=1,2} (\pm 1)^{a+b} \nonumber \\
&& \times \int d^3{\bf r}\, \psi_n^{(a) }({\bf r},t_f)\psi_n^{(b)*}({\bf r},t_f), 
\end{eqnarray}
such that the visibility and phase are given by the corresponding absolute value and phase of the mixed terms. We then have
\be Ve^{-i\delta\varphi }=\sum_n W_n\int d^3{\bf r}\, \psi_n^{(1)}({\bf r},t_f)\psi_n^{(2)*}({\bf r},t_f). \label{eq:Vis_n} \ee
First, note that the evolution of the center coordinates for each internal state $|1\rangle$ or $|2\rangle$ is independent of the initial wave function $\Phi_n$ In the trap. In addition, the scaling factors $\lambda_j$ are the same for all $\Phi_n$ if $\Phi_n$ are eigenstates of the initial harmonic trap. Under our assumption that the scaling factors are the same for the two arms it follows that in the case of a mixed-state input Eq.~(\ref{eq:overlap}) is generalized to 
\be V=\int d^3{\bf x}\,e^{-i\sum_j \sqrt{(\delta\bar{P}_j/\hbar)^2+( \delta X_j/2\sigma_j^2)^2}x_j}\rho_0({\bf x},{\bf x}), \ee
where $\rho_0({\bf r},{\bf r})$ is the initial atomic density in the trap. 
If initially the atoms in the trap are in a thermal state with temperature $T$ high enough so that the distribution is classical (a Boltzmann distribution) then the cloud has a Gaussian shape $\rho_0({\bf x})\propto \exp[-\sum_j x_j^2/2\Delta_j^2]$ with $\Delta_j=\sqrt{k_B T/m}/\omega_j$. The visibility is then
\be V=\exp\left[-\frac12\sum_{j=1}^3\left[(\Delta_j\delta\bar{P}/\hbar)^2+(\delta X_j/l_j)^2\right]\right],
\label{eq:overlap_thermal} \ee
where $l_j=2\sigma_j^2/\Delta_j=\hbar/\Delta p_j$ is a coherence length equal to the inverse of the momentum width of the atomic cloud $\Delta p_j=\sqrt{mk_B T}$.

\begin{figure}[t!]
\includegraphics[width=\columnwidth]{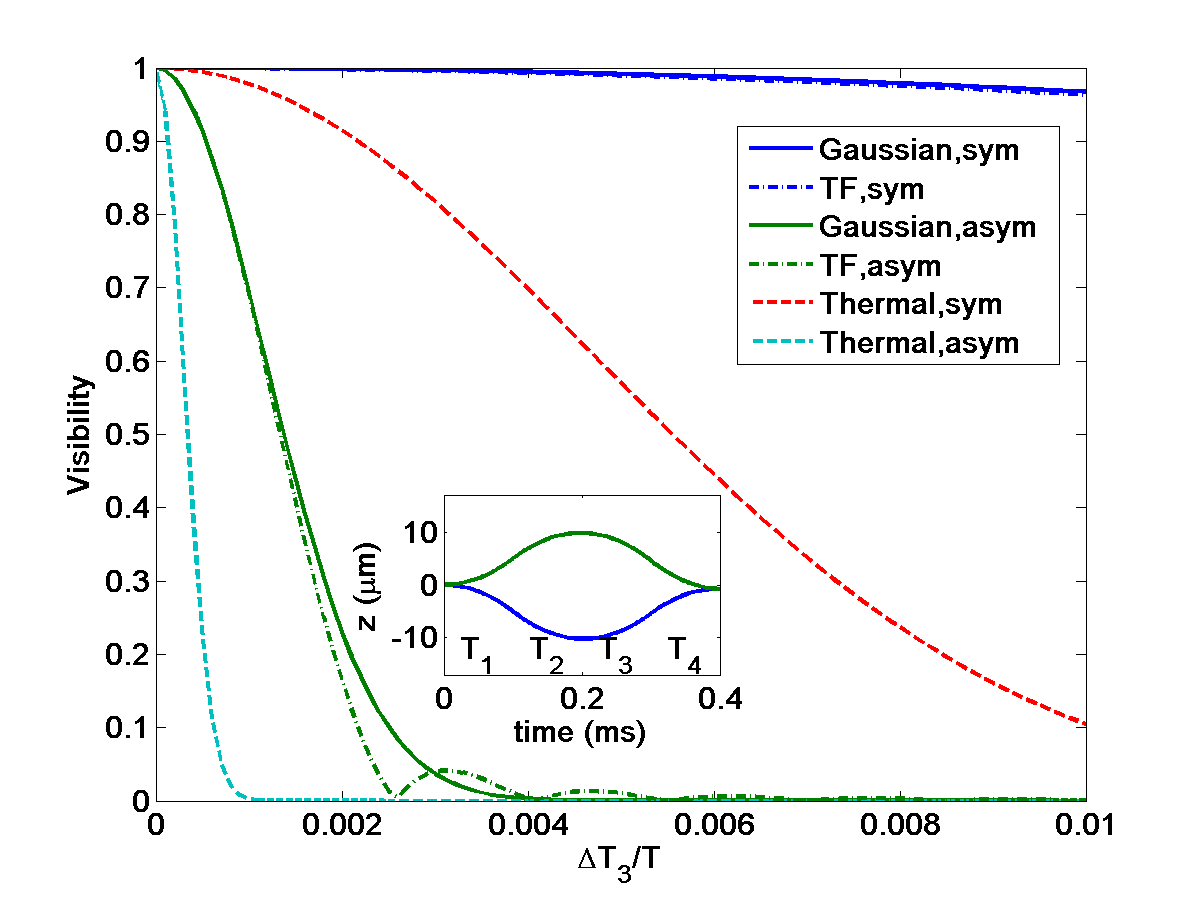} 
\caption{Spin coherence of a Stern-Gerlach interferometer as a function of recombination imperfections. $N=10^4$ atoms are prepared in a trap with the same parameters as in Fig.~\ref{fig:genTF} (longitudinal axis along $\hat{x}$). 
A $\pi/2$ pulse puts the atoms in an equal superposition of spin states 1\,ms after trap release and a sequence of 4 magnetic gradient pulses of equal durations $T=0.1$\,ms in the $\hat{z}$ direction (parallel to gravity) impose opposite accelerations $\pm a=\pm 10^3$\ m/s$^2$ for the two spin states. The gradient pulses (the second and third opposite to the first and fourth) split the two spin states into two counter-propagating wave-packets, stop their relative motion, accelerate them back towards each other and then stop their relative motion again. 
Ideally the trajectories of the two arms (see inset) terminate at the same position and with the same momentum, but a symmetric change in the durations of the two last pulses ($\Delta T_3=\Delta T_4$) causes a final relative spatial displacement $\delta Z\approx 4a T\delta T_3$ with zero relative momentum difference ($\delta P=0$), while an anti-symmetric change ($\Delta T_3=-\Delta T_4$) causes a momentum displacement  $\delta P=4ma\delta T_3$ as well, leading to a more drastic reduction of visibility. 
Thermal atoms at 100\,nK (dashed lines, cloud size $\Delta_z=12.4\,\mu$m, coherence length $l_z=94$\,nm) are more sensitive to imperfections than a BEC [solid lines for a Gaussian approximation, Eq.~(\ref{eq:overlap_pure}), dashed-dotted for a TF inverted parabolic shape, direct integration of Eq.~(\ref{eq:overlap1}), $\sigma_z=1.33\,\mu$m). By the time of interrogation (about 1.4\,ms after trap release) the cloud expands by a factor $\lambda_z\approx 1.33$ and the rate of expansion $\dot{\lambda}_z\approx 0.4$\,ms$^{-1}$ in Eq.~(\ref{eq:bardP}) significantly mixes the position and momentum terms in Eqs.~(\ref{eq:overlap_pure}) and~(\ref{eq:overlap_thermal}), such that the assumptions of the old theory of spin decoherence~\cite{Englert1988,Schwinger1988,Scully1989} become quantitatively invalid.}
\label{fig:HDinterf}
\end{figure}
 
In Fig.~\ref{fig:HDinterf} we use the above equations for calculating the spin coherence (visibility) of a Stern-Gerlach interferometer of the same type as proposed in the original work about the ``Humpty-Dumpty effect"~\cite{ Englert1988,Schwinger1988,Scully1989} and implemented recently in the lab~\cite{Amit2019}. We examine the drop of visibility when the interferometer uses either a BEC or thermal atoms and the recombination is not perfect. The interferometer sequence consists of four consecutive gradient pulses of equal strength and duration $T$ (see caption for parameters). A perturbation $\delta T$ in the duration of the last two pulses leads to predominantly a position displacement $\delta Z$ between the two arms at the output port if the perturbation is symmetric $T_3=T_4\to T+\delta T$,
and leads to a momentum displacement $\delta P_z$ if it is  antisymmetric $T_3\to T+\delta T$, $T_4\to T-\delta T$. 
For this interferometer sequence the position displacement for a symmetric perturbation is $\delta Z^{\rm sym}\sim 2\Delta aT\delta T$, where $\Delta a$ is the differential acceleration in each pulse, while the momentum displacement due to antisymmetric perturbation is $\delta P_z^{\rm anti}\sim 2m\Delta a\delta T$. It follows that $m\nu \delta X_z ^{\rm sym}/\delta \bar{P}_z^{\rm anti}\approx \nu T/\lambda^2$, where $\lambda^2\sim 1+\omega_z^2 T_f^2$ if the sequence is performed during expansion in free space ($\omega_z$ being the trapping frequency along the splitting direction and $T_f$ the total time of flight after trap release). 
This implies that perturbations in this kind of interferometer sequence lead predominantly to momentum displacement at the output port and this kind of imperfection plays the major role in reducing the interferometric visibility, as demonstrated in Fig.~\ref{fig:HDinterf}. 
 
The example presented in Fig.~\ref{fig:HDinterf} does not demonstrate the full novelty of the theory presented here and similar results could be obtained by methods presented in Ref.~\cite{Roura2014}, which treat either Gaussian wave-packets or BEC in the time-dependent TF approximation. These approximations for the thermal state or a BEC, respectively, are quite suitable for the present example. However, we emphasize that our treatment has the advantage that it unifies both cases into the same formalism and enables practical and easy predictions for the intermediate case where the atomic cloud does not satisfy the TF approximation. In addition, we provide a simple general expression [Eq.~(\ref{eq:overlap_thermal})] that permits an estimation of interferometric contrast based on the phenomenological lengths of cloud size and coherence length.

\section{Phase diffusion of propagating wave-packets}
\label{sec:phasediffusion}
 
As soon as an atomic BEC was realized and first exhibited an interference signal~\cite{Andrews1997}, it  became clear that its coherence is not limited
only by external noise after splitting, but also by intrinsic dynamics related to atom-atom interactions~\cite{ LewensteinYou1996,JavanainenYoo1996,CastinDalibard1997}. 
Theoretical studies of these dynamics often concentrate on a two-mode quantized model which reduces, under some assumptions, into the equivalent of a Josephson junction. These models give rise to tunneling oscillations between condensates~\cite{Smerzi1997,Zapata1998,Raghavan1999,Ostrovskaya2000,Anglin2001,Esteve2008,Gati2007,Giovanazzi2008,Ferrini2008} and to dephasing dynamics~\cite{ LewensteinYou1996,JavanainenWilkens1997,CastinDalibard1997,Pitaevskii2001,Boukobza2009,Grond2010,Fallen2015}. 
Some of the work in this field has also attempted to calculate the parameters of the two-mode models from first principles~\cite{JavanainenWilkens1997,Pitaevskii2001,Ananikan2006,Japha2011}, but they all use the spatially adiabatic approximation, where the instantaneous spatial modes are the steady-state solutions of the Gross-Pitaevskii equation in the respective potentials. The spatially non-adiabatic situation in which the two spatial modes of the BEC after splitting evolve in time or even propagate away from the splitting point has not been treated sufficiently and therefore deserves special consideration. 
 
Regardless of whether the splitting is spatially adiabatic or not, the dynamical evolution of the many-particle state is governed by a time scale of its own. Very slow splitting leads to a number-squeezed state where the uncertainty of the number difference between the two arms is sub-Poissonian and the phase uncertainty is large. For fast splitting, the number uncertainty is Poissonian while the phase uncertainty is relatively small. However, the number-uncertainty together with a number-dependent phase evolution due to atom-atom interactions lead to a dephasing effect that reduces the interferometric visibility~\cite{ LewensteinYou1996,JavanainenWilkens1997,CastinDalibard1997,Jo2007,Boukobza2009,Grond2010,Berada2013,Fallen2015}. Splitting at an intermediate rate leads to number squeezing with relatively slow dephasing~\cite{Jo2007,Esteve2008,Ferrini2008}. 
In contrast to decoherence due to random classical or quantum noise, phase diffusion due to interactions is in principle reversible, as shown experimentally in a Ramsey interferometer without spatial splitting~\cite{Widera2008}. 
 
In this section we develop a many-particle theory of BEC interferometry where the spatial dynamics may be non-adiabatic. The BEC wave-packets in the two interferometer arms follow the evolution presented in Sec.~\ref{sec:wpprop} with time-dependent parameters that also depend on the number of atoms in each arm. We take into account not only the central phase of each BEC wave-packet, which is given above in Eq.~(\ref{eq:varphi}), but also spatial features of the 
wave-packets. We find that these features become crucial for the interferometric visibility with guided matter waves. 
The theory is applied to interferometry with trapped, freely propagating or guided atoms and is valid for any number of atoms and strength of atom-aom interaction. 
 
\subsection{Evolution of the many-particle state} 
 
Consider coherent splitting of an $N$-particle BEC into two interferometer arms: ``left'' and ``right",  with corresponding single-particle wave functions $\psi_L({\bf r},0)$ and $\psi_R({\bf r},0)$ just after splitting (time $t=0$). 
If the two arms are fully separated then the subsequent evolution of each wave-packet depends on the number of particles in the corresponding arm, but not in the other one. 
In general, the many-particle wave function has the form
\be \Psi({\bf r}_1,\dots,{\bf r}_N,t)=\sum_{k=0}^N c_k \Psi_{k,N-k}({\bf r}_1,\dots,{\bf r}_N,t), 
\label{eq:manybody} \ee
where $c_k$ are the amplitudes for configurations with $k$ particles in the left arm and $N-k$ particles in the right arm, 
and the many-particle wave function of each configuration $\Psi_{k,N-k}$ evolves according to a many-particle Hamiltonian for a system with a fixed number of particles in each arm. If the particle-particle interactions are not too strong then the condensate approximation is valid, namely for a given number of particles we can assume that almost all the particles occupy the same spatial state. We can then use a mean-field approximation where the configuration wave function is
a symmetrized direct product of $k$ single-particle ``left'' wave functions and $N-k$ ``right'' wave functions
\begin{eqnarray} 
\Psi_{k,N-k}&\approx & e^{i\delta\chi_{k,N-k}(t)}
\hat{\cal S}\left\{ \psi_L^{(k)}({\bf r}_1,t)\dots \psi_L^{(k)}({\bf r}_k,t)\times \right. \nonumber \\
&& \left. \times\psi_R^{(N-k)} ({\bf r}_{k+1},t)\dots \psi_R^{(N-k)} ({\bf r}_N,t) \right\}. \label{eq:config} 
\end{eqnarray}
Here $\delta\chi_{k,N-k}(t)$ is a global phase that will be discussed below, $\hat{\cal S}$ is a symmetrization operator for bosons and the wave functions $\psi_L^{(k)}({\bf r},t)$ and $\psi_R^{(N-k)} ({\bf r},t)$ are the solutions of the time-dependent corresponding GPE,
\be i\hbar\frac{\partial\psi_a^{(n)}}{\partial t}=\hat{H}_{MF}^{(a,n)}\psi_a^{(n)}({\bf r},t), \label{eq:GP} \ee
where 
\be \hat{H}_{MF}^{(a,n)} = -\frac{\hbar^2}{2m}\nabla^2+V_a({\bf r},t)+gn|\psi_a^{(n)}({\bf r},t)|^2, \label{eq:H_MFan} \ee
and with initial conditions $\psi_a^{(n)}({\bf r},t=0)=\psi_a({\bf r},0)$ that are assumed to be independent of the number of particles. 
Here and in what follows $n$ and $a$ are dummy indices representing the number of particles and the interferometer arm labels, respectively: 
$n=k$ for $a=L$ and $n=N-k$ for $a=R$. 
In Eq.~(\ref{eq:H_MFan}) $H^{(a)}_0 \equiv -\hbar^2\nabla^2/2m+V_a$ includes the single-particle kinetic and potential energies, 
which may be different for left and right particles, and the last term is due to the mean-field repulsive atom-atom potential, as in Eq.~(\ref{eq:GPE0}). Note that here we assume that $k$ and $N-k$ are large numbers so that we will not be strict about whether the interaction in Eq.~(\ref{eq:H_MFan}) is proportional to $n$ or $n-1$. 
 
Here we focus on the long-time evolution after splitting. Details of the splitting process itself determine the coefficients $c_k$, which evolve during  the time where two arms are not yet fully separated and the evolution is not fully described by Eq.~(\ref{eq:GP}). In a quick momentum splitting by a Bragg or Raman process, the coefficients $c_k$ are almost fully determined when the two arms separate in momentum space: $\psi_{L/R}({\bf r},0)\approx \psi({\bf r},t<0)e^{ i{\bf K}_{L/R}\cdot{\bf r}}$, where $\psi({\bf r},t<0)$ is the wave-packet of the BEC before splitting and ${\bf K}_{L/R}$ are the momentum kicks corresponding to the two arms. 
Then the two wave-packets separate quickly in space so that the time of interaction between them is too short to affect the number distribution. The configuration amplitudes $c_k$  in Eq.~(\ref{eq:manybody}) then represent a binomial number distribution
\be |c_k|^2\equiv {\cal P}_k=\frac{1}{2^{N}}\left(\begin{array}{c} N \\ k \end{array}\right) 
\to \frac{ e^{-(k-N/2)^2/2\Delta n^2}}{\sqrt{2\pi}\Delta n}
\label{eq:c_k} \ee
of width $\Delta n\approx \sqrt{N}/2$ around $k=N/2$. 
In contrast, if the splitting is slow then number squeezing leads to a final number distribution that is narrower than Poissonian. 
 
The additional configuration phase $\delta\chi_{k,N-k}(t)$ in Eq.~(\ref{eq:config}) is a residual phase added to the sum of global single-particle phases $k\varphi_L^{(k)}+(N-k)\varphi_R^{(N-k)}$ of the single-particle wave functions. 
$\delta\chi_{k,N-k}$ can be derived from the Schr\"odinger equation for the many-body state $i\hbar\partial\Psi/\partial t=\hat{\cal H}_N\Psi$, where the many-particle Hamiltonian is 
\be \hat{\cal H}_N=\sum_{p=1}^N\left[-\frac{\hbar^2\nabla_p^2}{2m}+\hat{V}({\bf r}_p,t)+\sum_{q<p} U({\bf r}_p-{\bf r}_q)\right],
\label{eq:Hmb} \ee
where  the external potential $\hat{V}$ may depend on internal degrees of freedom, if attached to the interferometer arms, and $U({\bf r} -{\bf r}')$ is the inter-particle potential, which we usually approximate for slow collisions by $U({\bf r} -{\bf r}')=g\delta({\bf r} -{\bf r}')$. If we assume no overlap between $\psi_L$ and $\psi_R$ then the Schr\"odinger equation separates into independent equations for the two arms, such that
\be \delta\chi_{k,N-k}=\delta\chi_L^{(k)}+\delta\chi_R^{(N-k)}. \ee
By using the GPE in Eq.~(\ref{eq:GP}) and integrating over all coordinates we then obtain for each of these residual phases
 
\be  -\hbar\frac{\partial\delta\chi_a^{(n)}}{\partial t}=
\langle\langle \hat{\cal H}_N\rangle\rangle_a
-n \langle \hat{H}_{MF}^{(a,n)}\rangle,
\label{eq:Schroedinger} \ee
namely, the rate of change of the residual global phase of the $n$ particle system is the difference between the expectation value of the many-particle Hamiltonian and the sum of the expectation values of the single-particle mean-field Hamiltonian for all the particles. 
 
Note that in Eq.~(\ref{eq:Schroedinger}) the single-particle parts of the many-body Hamiltonian in Eq.~(\ref{eq:Hmb})
cancel with the collisionless terms of $\hat{H}_{MF}^{(a,n)}$ of Eq.~(\ref{eq:H_MFan}) and we are left with
\be \frac{\partial \delta\chi_a^{(n)}}{\partial t}= 
\frac{g}{\hbar}\frac{n^2}{2}\int d^3{\bf r}\,
\left|\psi_a^{(n)}\right|^4, 
\label{eq:dchikdt} \ee
which compensates for the double counting of pair interactions when summing up the single-particle energies. 
It follows that the rate of  change of the total global phase of each configuration that includes the sum of single-particle phases in Eq.~(\ref{eq:varphi}) is 
\be \dot{\chi}_a^{(n)}= -\frac{n}{\hbar}\left[\frac{5n}{7N}\frac{\mu_{\rm int}}{\prod_j\lambda_{j,a,n}}+\sum_{j=1}^3 \frac{\hbar\nu_j}{2\lambda_{j,a,n}^2}\right], 
\label{eq:dotchi_an} \ee
where we have used $gN\int d^3{\bf r}\, |\Phi_0({\bf r})|^4=E_{\rm int}=\frac47\mu_{\rm int}$ from Eq.~(\ref{eq:E_int}).

\subsection{Interferometric visibility}
 
In atom interferometry the relative phase between two spatial paths can be probed in different ways. If the atoms in the two arms are labeled by different internal (spin) states then the phase may be revealed by bringing the two wave-packets into full overlap (in both position and momentum) and then probing the internal state of the recombined wave-packet. If the two arms are not distinguishable by the internal state, then the phase may be revealed by spatial interference fringes formed when the two wave-packets overlap in space (but not in momentum) or alternatively by applying a Bragg sequence that transforms the spatial fringe pattern into momentum components with a probability dependent on the phase. 
For brevity we discuss here only spatial fringes as a probe of the phase and the results will apply with small modifications also to the alternative  interferometric schemes. 
We therefore assume that the wave-packets in the two arms have the same internal state and are distinguishable by their orthogonal spatial state $\int d^3{\bf r}\, \psi_L^*({\bf r},t)\psi_R({\bf r},t)=0$ at any time $t$ and specifically at the time where the fringe pattern is imaged.  
 
The atomic density follows from the many-particle wave function in Eqs.~(\ref{eq:manybody}) and~(\ref{eq:config}):
\begin{eqnarray}
 \rho({\bf r},t) &=& \int d^3{\bf r}_2\dots\int d^3{\bf r}_N\, |\Psi({\bf r},{\bf r}_2,\dots,{\bf r}_N,t)|^2 \nonumber \\ 
&=& \sum_{k=0}^N |c_k|^2\left(\frac{k}{N}|\psi_L^{(k)}|^2+\frac{N-k}{N}|\psi_R^{(N-k)}|^2\right)
\nonumber \\
&& +\sum_{k=1}^N\left(
c_k^*c_{k-1}\frac{\sqrt{k(N-k+1)}}{N}A_L^{(k)}A_R^{(N-k+1)*} \times \right. \nonumber \\
&&  \times e^{-i(\delta\chi_{k,N-k}-\delta\chi_{k-1,N-k+1})}\psi_L^{(k)*} ({\bf r},t)\psi_R^{(N-k+1)}({\bf r},t) \nonumber \\
&& \left. +{\rm c.c.}\right), 
\label{eq:rhor1}
\end{eqnarray}
where 
\be A_a^{(n)}=\left[\int d^3{\bf r}\, [\Phi_a^{(n)}({\bf r})]^*\Phi_a^{(n-1)}({\bf r})\right]^{n-1} 
\label{eq:ovlp} \ee
is the product of the overlap integrals of the single-atom wave functions of the same arm $a=L$ or $a=R$ with different mean-field potential due to different numbers of particles in the arm. For all number configurations the center coordinates for each arm are the same, so that we have expressed the overlap integral in terms of the wave-packet wave functions $\Phi_a$ in the frame moving with the external coordinates ${\bf R}_a$, as $\psi_a^{(n)}({\bf r})=\Phi_a^{(n)}({\bf r}-{\bf R}_a)e^{i{\bf P}_a\cdot({\bf r}-{\bf R}_a)/\hbar}$. 
It is easy to see that in the absence of interactions, where $\phi_L$ and $\phi_R$ are independent of the number of particles and $c_k$ have the binomial form of Eq.~(\ref{eq:c_k}), $A_a^{(n)}=1$ and the configuration phases $\delta\chi_{k,N-k}$ vanish, Eq.~(\ref{eq:rhor1}) reduces to the single-particle density $\rho({\bf r})=\frac12(|\psi_L+\psi_R|^2$. 
 
The overlap integrals $A^{(n)}_a$ in Eq.~(\ref{eq:ovlp}) have two effects on the interference term (bottom lines) in Eq.~(\ref{eq:rhor1}]. They add phase and may also reduce the amplitude of the interference term if $|A_a^{(n)}|<1$. In Appendix~\ref{app:number_phase} we show that the latter effect is negligible in most practical cases. We now assume that the wave functions in the center frames of the two arms  have the form of Eq.~(\ref{eq:Ansatz}) and that the number-dependence of the wave-packet envelopes plays a very minor role. 
The interference term in Eq.~(\ref{eq:rhor1}) may then be approximated by
\begin{eqnarray}
\rho_{ LR}&& = |\Phi_L({\bf r}-{\bf R}_L,t)|\Phi_R({\bf r}-{\bf R}_R,t)| e^{-i[\delta \tilde{S}+\delta{\bf P}\cdot({\bf r}-\bar{\bf R})]/\hbar}\times 
\nonumber \\
&& \times \sum_{k=1}^N c_k^*c_{k-1} \frac{\sqrt{k(N-k+1)}}{N}e^{-i[\phi_L^{(k)}-\phi_R^{(N-k+1)}] } \nonumber \\
&& \times e^{-\frac{i}{2}\sum_{j=1}^3 (\alpha_j^{(L,k)}(r_j-R_{Lj})^2-\alpha_j^{(R,N-k+1)}(r_j-R_{Rj})^2) }.
\label{eq:interf-term} 
\end{eqnarray}
Here the phase terms in the first line are related to the wave-packet trajectories and are therefore independent of the particle number: $\delta \tilde{S}=S_L-S_R-\bar{\bf P}\cdot\delta{\bf R}$ is the difference of actions along the two paths with a correction due to the difference of the path endpoints, $\bar{\bf R}$ and $\bar{\bf P}$ are the center-of-mass positions and momentum, respectively, and $\delta{\bf R}={\bf R}_L-{\bf R}_R$ and $\delta{\bf P}={\bf P}_L-{\bf P}_R$ are the corresponding differences  of the trajectory endpoints. 
The coordinate-independent phases in the second line of Eq.~(\ref{eq:interf-term}) are given by (see Appendix~\ref{app:number_phase})
\be \phi_a^{(n)}=\left.\frac{\partial\chi_a^{(n)}}{\partial n}\right|_{n-\frac12}
+(n-1)\sum_{j=1}^3 \frac{\lambda_{j,a,n}\lambda_{j,a,n-1}}{4\nu_j}
\left. \frac{\partial \xi_j^{(a,n)}}{\partial n}\right|_{n-\frac12}
\label{eq:phi_a_n} \ee
which involves the derivatives with respect to the particle number (at $n-\frac12$) of the global configuration phases $\chi_L$ and $\chi_R$ given by the time integral over Eq.~(\ref{eq:dotchi_an}) and of the relative expansion rate along the axes $j$ of each wave-packet
\be \xi_j^{(a,n)}\equiv \frac{\dot{\lambda}_{j,a,n }}{\lambda_{j,a,n }}=\frac{\hbar}{m}\alpha_j^{(a,n)}, \label{eq:xi} \ee
with the scaling factors $\lambda_{j,a,n}$ for each arm and particle number being the solutions of Eq.~(\ref{eq:ddlambda}) with $\eta_a=n/N$. 
The momentum chirp coefficients $\alpha_j^{(a,n)}$ relative to the corresponding expansion rate coefficients $\xi_j^{(a,n)}$ [see also Eq.~(\ref{eq:alpha})] appear in the 
third line of Eq.~(\ref{eq:interf-term}) in the coordinate-dependent phase responsible for the formation of the spatial interference fringes in the case of two distant wave-packets at relative rest ($\delta{\bf P}=0$) expanding into each other and overlapping in space (the atomic analogue of  a double-slit experiment). However, in interferometer schemes that use two internal atomic states as the two arms or a Bragg sequence to extract the interferometric phase at the output port the term at the third line is integrated over and may give rise to the reduction of visibility if the final wave-packet positions ${\bf R}_L$ and ${\bf R}_R$ do not overlap or if the expansion coefficients are different for the two arms.

Let us now assume that the average number of atoms in the two arms is equal, such that $\sum_{k=0}^N |c_k|^2 k=N/2$. Under the same assumption that led to Eq.~(\ref{eq:interf-term}) the wave-packet envelopes do not depend on the particle numbers $k,N-k$ and the atomic density of Eq.~(\ref{eq:rhor1}) becomes
\begin{eqnarray} \rho({\bf r}) &=& \frac12\left[|\psi_L|({\bf r})^2+|\psi_R({\bf r})|^2 \right. \nonumber \\
&&\left. +2|\psi_L({\bf r})||\psi_R({\bf r})|C\cos\delta\bar{\phi}({\bf r})\right]. 
F
\end{eqnarray}
Here $|\psi_a({\bf r})|=|\Phi_a({\bf r}-{\bf R}_a)|$
and the phase $\delta\phi({\bf r})$ is given by
\be \delta\bar{\phi}({\bf r})=\left. \delta\phi_k({\bf r})\right|_{k=\frac12(N+1) }, \ee
where 
\begin{eqnarray}  
&&\delta\phi _k ({\bf r}) = \frac{1}{\hbar}\left[\delta\tilde{S}+\delta P\cdot({\bf r}-\bar{\bf R})\right]+
\phi_L^{(k)}-\phi_R^{(N-k+1)} \label{eq:deltaphir}
\\
&& +\frac{1}{2}\sum_j \left[\alpha_j^{(L,k)}(r_j-R_{Lj})^2 -\alpha_j^{(R,N-k+1)}(r_j-R_{Rj})^2\right] . \nonumber
\end{eqnarray}
This expression for the phase is based on the assumption that the distribution ${\cal P}_k=|c_k|^2$ is symmetric around $k=N/2$ and that the coefficients $c_k$ are real. 
The contrast $C$ is obtained by summation over the different particle numbers
\be C=\sum_k c_k^*c_{k-1}e^{-i(\delta\phi_k-\delta\bar{\phi})}, \ee
which is real if the $c_k$ are symmetric around $k=N/2$. 
By taking the coefficients $c_k$ to be approximated by a symmetric Gaussian distribution as in Eq.~(\ref{eq:c_k}) with arbitrary distribution width $\Delta n$ and expanding the deviation of the phase linearly around $N/2+\frac12$ we obtain
\begin{eqnarray}
 C &=& \sum_k c_k c_{k-1}
\exp\left[i\left(k-\frac{N}{2}-\frac12\right)\left.\frac{\partial\delta\phi_k}{\partial k}\right|_+{k=\frac12(N+1)}\right] \nonumber \\ 
&\approx & 
\exp\left\{-\frac12 \left[\Delta\phi_0^2+\Delta \phi_t({\bf r})^2\right]\right\},  \label{eq:pdcontrast}
\end{eqnarray}
where $\Delta\phi_0=1/2\Delta n$ is the initial phase uncertainty due to the number uncertainty $\Delta n$ just after splitting, and 
\be \Delta\phi_t({\bf r})\equiv  \Delta n\left. \frac{\partial \delta\phi_k({\bf r})}{\partial k}\right|_{k=\frac12(N+1)}. \ee
In general, the phase $\delta\phi_t$ and hence contrast $C$ are coordinate dependent.  However, if the expansion coefficients $\alpha_j^{(a,n)}$ are not too large and not much different for the two arms we may neglect this dependence and look only at the visibility in the middle of the interference fringe at ${\bf r}=\bar{\bf R}$. We then neglect the second line of  Eq.~(\ref{eq:deltaphir}) and obtain
\be \Delta\phi_t\approx \Delta n\left. \frac{\partial}{\partial k} [\phi_L^{(k)}-\phi_R^{( N-k+1)}]\right|_{k= \frac12(N+1)} =\Delta\phi_L+\Delta\phi_R, \ee
where, upon using Eq.~(\ref{eq:phi_a_n}), 
\be \Delta\phi_a \approx \Delta n\left[\frac{\partial}{\partial n}\left(
\frac{\partial\chi_a^{(n)}}{\partial n}+n\sum_j\frac{\lambda_{j,a,n}^2}{4\nu_j}\frac{\partial\xi_j^{(a,n)}}{\partial n}\right)\right]_{n=N/2}.
\label{eq:dphi_tot} \ee
By using the explicit expression for $\dot{\chi}_a^{(n)}$ in Eq.~(\ref{eq:dotchi_an})  and replacing $n\to N\eta$, where $\eta=\eta_a$ is the fraction of atoms in the corresponding arm, we find
\begin{eqnarray} \Delta\phi_a &=& 
\frac{\Delta n}{N}\left[-\int_0^t dt'\,\frac{\partial^2}{\partial\eta^2} 
\left(\frac{5\mu_{\rm int} }{7\hbar}\frac{\eta^2}{\Lambda_a^{(n)}}
+\sum_{j=1}^3 \frac{\nu_j}{2}\frac{\eta}{\lambda_{j,a,n}^2}\right)\right.
\nonumber \\
&&\left.+ \sum_{j=1}^3 \frac{\partial}{\partial\eta}\left(\frac{\eta \lambda_{j,a,n}^2}{4\nu_j}\frac{\partial\xi_j^{(n)}}{\partial\eta}\right)\right],
\label{eq:dphi_tot1}
\end{eqnarray}
where the derivatives with respect to the number fraction $\eta$ are taken at $\eta=1/2$ and $\Lambda_a^{(n)}=\lambda_{1,a,n}\lambda_{2,j,n}\lambda_{3,j,n}$ is the relative wave-packet volume at a given time. 
 
The time evolution of the phase uncertainty $\Delta\phi_t$ is proportional to the number uncertainty $\Delta n$ during the splitting and determined by the number dependence of three properties of the wave-packets: (i) the interaction energy, proportional to $\eta^2\mu_{\rm int}/\Lambda_a(t)$, (ii) the kinetic energy $\hbar\nu_j/\lambda_j^2$, determined by the wave-packet width, and (iii) the expansion rates $\xi_j$, which are determined by the wave-packet evolution at short times after release from a trap. 
In the following we will compare our result to previous results concerning spatially adiabatic splitting in a double-well potential and obtain new results for cases where the atomic wave-packets are expanding during propagation along the interferometer arms, where the term in the bottom line of Eq.~(\ref{eq:dphi_tot1}) becomes important. 
\subsection{Application to specific schemes}

\subsubsection{Spatially adiabatic splitting}

In spatially adiabatic splitting, the single-particle wave function of the BEC follows the stationary solution of the GPE [Eq.~(\ref{eq:GPE0})] for the instantaneous potential and number of particles in each arm. The wave function sizes $\sigma_j(t)=\lambda_j\sigma_j(0)$ in the two arms, with particle numbers $\eta_LN$ and $\eta_R N$, satisfy the stationary state equation [Eq.~(\ref{eq:selfcons})] with $N\to \eta_aN$ ($\eta_L+\eta_R=1$). 
Equivalently, the scaling factors $\lambda_j$ satisfy Eq.~(\ref{eq:ddlambda}) with $\ddot{\lambda}_j\to 0$ and the expansion rates $\xi_j$ [Eq.~(\ref{eq:xi})] vanish. 
 
Let us consider a symmetric double-well potential where after splitting the atoms in the two arms reside in two potential wells with the same frequencies $\omega_j$. 
In this case the rate of change of the configuration phases $\chi_a^{(n)}$ of Eq.~(\ref{eq:dotchi_an}) for any particle number are nothing but the configuration energies, and their derivatives with respect to the particle numbers in Eq.~(\ref{eq:phi_a_n}) are the chemical potentials $\partial \chi_a^{(n)}/\partial n=\eta_a\mu_{\rm int}/\Lambda_a+\frac12\hbar\sum_j\nu_j/\lambda_{j,a,n}^2$, while the second term in Eq.~(\ref{eq:phi_a_n}) that involves the expansion rates vanishes. 
From the stationary limit of Eq.~(\ref{eq:ddlambda}) for the scaling factors, it follows that in the TF limit $\lambda_j(\eta)=\eta^{1/5}$ (we may assume that the trap frequencies at time $t$ are equal to those of the initial trap, since the history in the adiabatic limit is not important). 
It follows that $\eta^2/\Lambda=\eta^{7/5}$ in the first term of the integrand of Eq.~(\ref{eq:dphi_tot1}) and the second term is negligible, so that the phase diffusion rate becomes
     \be \Gamma\equiv \frac{\partial \Delta\phi_t}{\partial t}= \frac{8}{5}\frac{\Delta n}{N}\frac{\mu_{\rm int}(N/2)}{\hbar}, \ee
where $\mu_{\rm int}(N/2)=\frac12\mu_{\rm int}(N)/\Lambda$ is the interaction chemical potential for a system of $N/2$ particles. 
This result coincides with previous predictions of the phase diffusion rate of split condensates~\cite{ JavanainenWilkens1997,CastinDalibard1997,PDremark}. 
If the number distribution between the interferometer arms is Poissonian with $\Delta n=\sqrt{N}/2$ then the phase diffusion rate scales as $\Gamma(N)\propto N^{2/5}/\sqrt{N}=N^{-1/10}$.  This is a rather weak dependence on the particle number but for small $N$ this result is non-physical, since we should expect the effect of phase diffusion to vanish in the limit of a very dilute Bose gas with negligible atom-atom interactions. 
 
\begin{figure}[t!]
\includegraphics[width=\columnwidth]{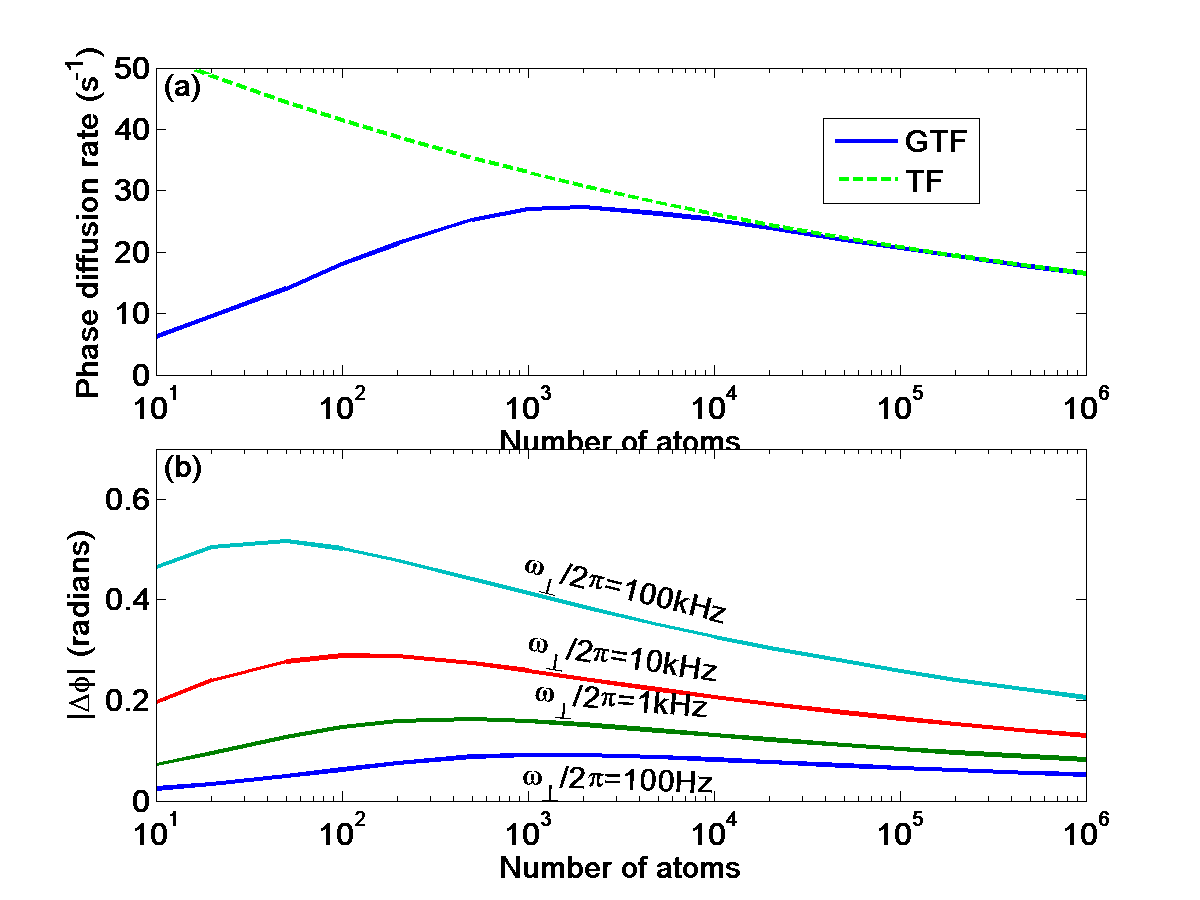} 
\caption{Phase diffusion as a function of the number of particles for a Poissonian number distribution $\Delta n=\sqrt{N/2}$. 
(a) Phase diffusion rate in a double-well potential after spatially adiabatic splitting. Each of the wells has trap parameters as in Fig.~\ref{fig:genTF}. The Thomas-Fermi (TF) approximation for the phase diffusion rate $\Gamma\propto N^{-1/10}$ (dashed curve) is compared to the GTF result (solid curve), which drops for low atomic densities.
(b) Asymptotic value of the phase uncertainty for a freely expanding split BEC after release from an initial trap with transverse frequency $\omega_{\perp}$ ($\omega_{\parallel}/\omega_{\perp}=0.4$). Interference visibility declines significantly only for very tight traps and a small number of particles. The other parameters are for $^{87}$Rb as in all figures.}
\label{fig:pdrate}
\end{figure}
 
In Fig.~\ref{fig:pdrate}(a) we compare the phase diffusion rate for the TF approximation (dashed curve) to the more accurate prediction based on the generalized TF approximation of this work (solid curve). 
The latter prediction produces the expected behavior where the phase diffusion rate  decreases for small particle numbers while it coincides with the prediction of the TF approximation for large particle numbers. 
 
\subsubsection{Free expansion}
 
Consider a BEC released from a cylindrical trap with $\omega_{\perp}\gg \omega_{\parallel}$ at $t=0$ and immediately split by a quick pulse before expansion begins (at $t\ll \omega_{\perp}^{-1}$).  The evolution of the scaling factor $\lambda_{\perp}$ is then given by Eq.~(\ref{eq:lambdaperpt}) with $\lambda_0=1$ and $\dot{\lambda}_0=0$, where we assume that $\lambda_{\parallel}\sim 1$, namely
$\lambda_{\perp}(t)=\sqrt{1+(\eta\tilde{\omega}_{\perp}^2+\nu_{\perp}^2)T^2}$, where $\eta\sim \frac12$ and $\tilde{\omega}_{\perp}^2=\omega_{\perp}^2-\nu_{\perp}^2$. By using $\mu_{\rm int}=\frac72 m\tilde{\omega}_{\perp}^2\sigma_{\perp}^2=\frac74\hbar\tilde{\omega}_{\perp} /\nu_{\perp}$ the phase uncertainty becomes
\begin{eqnarray}
\Delta\phi_t &=& \frac{\Delta n}{N}
\frac{\tilde{\omega}_{\perp}^2}{\nu_{\perp}}\left[-\int_0^t \frac{dt'}{\lambda_{\perp}(t')^6}
(\nu_{\perp}^2t'^2+1)(\nu_{\perp}^2t'^2+5) \right. \nonumber \\
&&\left. + \frac{t}{\lambda_{\perp}(t)^4}(1+\nu_{\perp}^2 t^2)\right]
\end{eqnarray}
The long-time limit of this phase when $t\gg \sqrt{\nu_{\perp}^2+\tilde{\omega}_{\perp}^2/2}$ is 
\be \Delta\phi_{t\to \infty}=-\frac{3\pi\Delta n}{8 N}\frac{b}{\sqrt{1+b}}(5+\frac{2}{1+b}+\frac{1}{(1+b)^2}), 
\ee
where $b=\tilde{\omega}_{\perp}^2/2\nu_{\perp}^2$. In the TF limit $\tilde{\omega}/\nu_{\perp}\to \frac{4}{7}\mu_{\rm int}/\hbar\omega_{\perp}$ and $b\gg 1$, so that for a Poissonian distribution $\Delta n=\sqrt{N/2}$ we obtain
\be \Delta\phi_{t\to\infty} \to \frac{15\pi \mu_{\rm int} }{28\hbar\omega_{\perp}\sqrt{N}}
\approx \frac{15\pi}{28 N^{1/10}}\left(\frac{15}{8}\frac{\omega_{\parallel}}{\omega_{\perp}}\frac{a_s}{\ell_{\perp}}\right)^{2/5}.  \ee
In Fig.~\ref{fig:pdrate}(b) we present the asymptotic limit of the phase uncertainty due to phase diffusion for free expansion as a function of the number of particles $N$ and for different values of $\omega_{\perp}$. For large particle numbers the $N^{-1/10}$ behavior dominates for all values of the trap frequency. However, when the number of particles becomes smaller the phase uncertainty reaches a maximum which, for typical trapping frequencies, is still quite low and generates little reduction of interference visibility. We therefore conclude that phase diffusion due to atom-atom interactions does not represent a major limitation for interferometry with freely propagating atoms. 
 
\subsubsection{Expansion in a waveguide}
\

Although quite a few proposals and much experimental effort has been devoted to guided matter-wave interferometry with a promise for highly compact and accurate inertial sensing~\cite{Gupta2005,Wu2007,Japha2007,Baker2009,Sherlock2011,Turpin2015,Navez2016,Pandey2019}, coherence signals from such a device have not yet been demonstrated. Difficulties may lie in the implementation of smooth enough magnetic or optical potentials for such waveguides -- for example, in a loop configuration for Sagnac interferometry -- but one of the main obstacles, even if such potentials may be constructed, is that a pair of coherently split BEC wave-packets kept confined in a waveguide for a long time must suffer from phase diffusion due to atom-atom interactions. 
\begin{figure}[t!]
\includegraphics[width=\columnwidth]{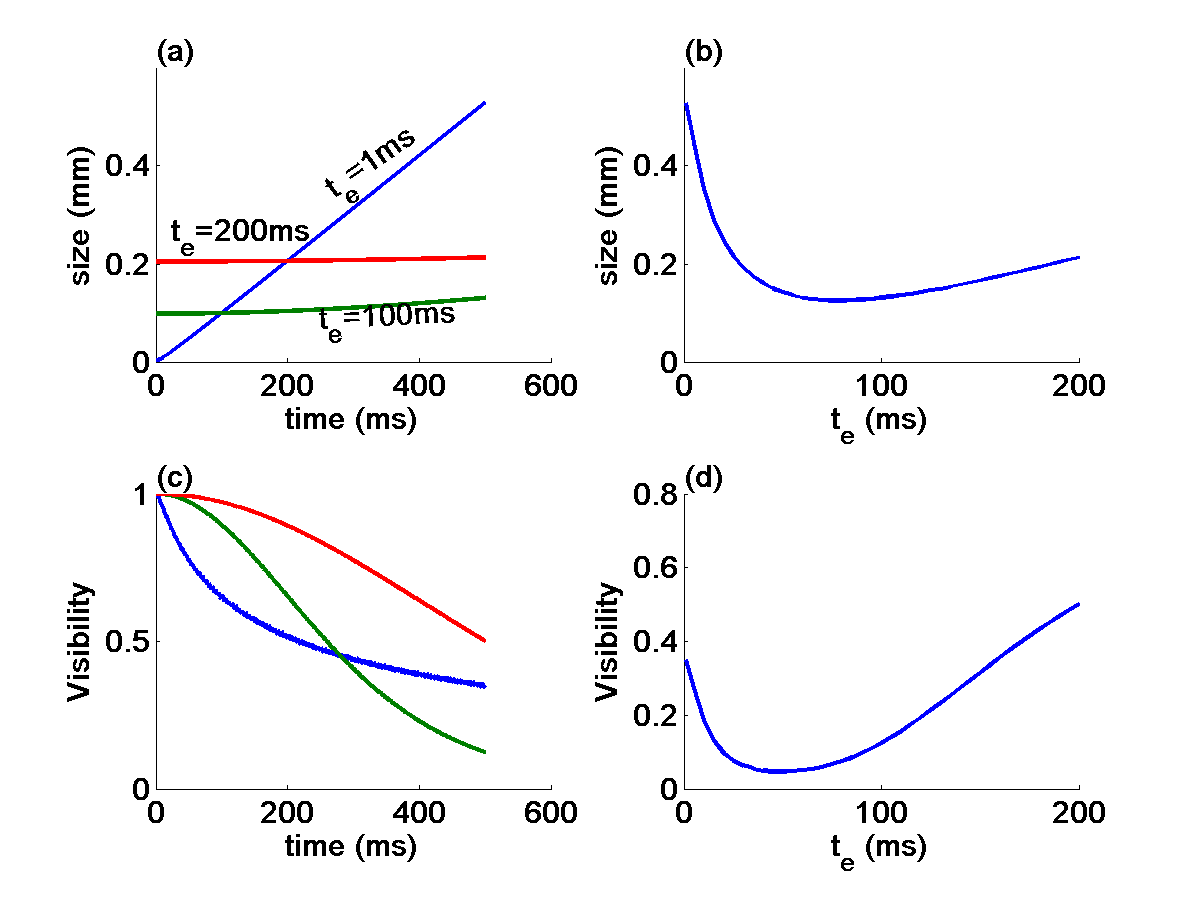} 
\caption{Phase diffusion of guided matter waves.  A BEC of $10^4$ $^{87}$Rb atoms in a trap with the same parameters as in Fig.~\ref{fig:genTF} is released into a waveguide with 
the same transverse frequency as the initial trap, $\omega_{\perp}=2\pi\times 100$\,Hz. 
The longitudinal trap potential is ramped down linearly within $\tau=1$\,ms and the BEC is allowed to expand along the waveguide for a time $t_e$, after which the longitudinal harmonic potential is applied again as a collimation pulse for a duration $T_{\rm col}=\dot{\lambda}_x/\lambda_x\omega_x^2$ (a method known as ``delta-kick collimation"~\cite{Hannover2012,Abend2016}).
These steps are completed  before splitting. The evolution of the cloud size [in (a)] and the coherence [in (c)] depends on the initial expansion time $t_e$. If $t_e$ is small then the collimation is not effective and the cloud continues to expand, while the coherence decreases due to phase diffusion. For intermediate expansion times ($t_e=100$\,ms) the final cloud is relatively small and dense, giving rise to enhanced phase diffusion after a long time of propagation, while only long initial expansion times $t_e$ give rise to a BEC cloud with a stable size large enough to slow down the rate of phase diffusion. Panels (b) and (d) show the cloud size and coherence after a propagation time of 0.5\,sec after splitting as a function of $t_e$, which demonstrates the advantage of a long collimation process giving rise to a medium-sized cloud and fairly high coherence.}
\label{fig:guiderelease}
\end{figure}
 
Fig.~\ref{fig:guiderelease} demonstrates the application of the wave-packet evolution theory and the phase diffusion theory of this section for optimizing the performance of guided matter-wave interferometers. We consider a guided interferometer based on a trapped BEC released into free expansion in the waveguide and then split into two counter-propagating wave-packets that are brought together again after a time $t$ (for example, in a ring configuration, where the curvature of the guiding potential is neglected here for simplicity). 
We wish to keep the size of the clouds propagating along the waveguide as small as possible for easier manipulation but on the other hand prevent the harmful effect of high density on the coherence. 
We therefore perform a collimation procedure before splitting, consisting of free expansion for a time $t_e$ followed by ramping up the longitudinal harmonic potential for a time $T_{\rm col}=\dot{\lambda}_x/\lambda_x\omega_x^2$. In Fig.~\ref{fig:guiderelease} we show the cloud size after splitting and the evolution of the contrast as a function of time after this collimation process and splitting. 
If the cloud size is not too small the phase diffusion rate may be 
reduced, thereby retaining fairly high contrast at the interferometer output after time $t$. 
 
The calculation of the coherence in Fig.~\ref{fig:guiderelease} is based on Eq.~(\ref{eq:dphi_tot1}), where symmetry between the arms is assumed, such that the evolution for a given number of particles is exactly the same in the two arms. 
For the range of parameters used here, the dominant contribution to the phase diffusion 
comes from the first term -- the self-interaction of the wave-packet, while the third term -- the number-dependent expansion phase -- has a maximal contribution 
of about 5\% to the decoherence. However, in general the new sources of intrinsic decoherence that were found here should be taken into account when developing new approaches to guided matter-wave interferometry. 
 
\section{Conclusions and outlook}
\label{sec:outlook}

The generalized wave-packet evolution theory presented here provides an efficient tool for calculating the performance of atomic interferometers based on trapped, guided or freely propagating atomic clouds. It is valid for atomic evolution in time-dependent potentials that are smooth on the scale of size of the atomic clouds and for atom-atom interactions ranging from negligible 
interactions (as for a single atom or a dilute thermal cloud) 
to a strongly interacting BEC, as long as the condensate approximation holds. This generalizes previous approaches that were valid for certain ranges of interactions or potentials~\cite{Jamison2011,Roura2014,Meister2017}. In particular, we first demonstrated the validity of our theory for static properties of a BEC in a cylindrical trap with a relatively low aspect ratio between the trap axes or a very high aspect ratio with a transition between a 3D and a quasi-1D BEC (Sec.~\ref{sec:GTF}) 
and then for dynamical evolution when a trapped BEC is released into free space or a matter waveguide (Sec.~\ref{sec:solutions}). The calculations based on the wave-packet evolution theory are compared to direct solutions of the GPE and excellent agreement is obtained over the whole range of atomic interactions. This gives us confidence that the theory is valid not only in the no-interaction or strong-interaction limits, where it converges with existing theories, but also over the intermediate regime where this work provides a unique 
method, heretofore unavailable for such a wide range of scenarios in the context of atom interferometry. 
 
The most important physical entity in an interferometer is the phase difference between 
different paths. However, in most interferometric situations it is not sufficient to follow the dynamical phase along a single trajectory 
in each interferometer arm. It is also necessary to follow the evolution of the whole wave-packet along the arms for a few reasons.  First, the internal wave-packet dynamics can contribute to the phase, as shown in Eq.~(\ref{eq:varphi}). Second, if the wave-packets in the two arms do not precisely overlap at the output port then the phase and the visibility of the interferometer is changed, as discussed thoroughly in Sec.~\ref{sec:spincoherence}. In addition, 
accurately including wave-packet dynamics is necessary to describe the evolution in open interferometers based on spatial interference fringes, as demonstrated in Ref.~\cite{Margalit2019}. In all these cases the theory presented in this paper is expected to provide the necessary tools for predicting interferometric performance. 
 
One novelty of this work that makes it most suitable for treating various interferometric scenarios is that it includes 
effects due to changing the number of atoms in a BEC wave-packet when an initial cloud is split into separate clouds, each including a fraction of the total number of atoms. One of the important consequences of the dependence of the evolution on atom number is phase diffusion due to atom-atom interactions, which arise from the number uncertainty after splitting. This effect is thoroughly discussed in Sec.~\ref{sec:phasediffusion}, which presents a theory of phase diffusion in dynamic interferometric situations that 
has not been treated before. 
 
For the situations discussed in this work, we stress that wave-packet dynamics could, in principle, be calculated by direct  numerical solution of the Schr\"odinger equation or the GPE. However, in most practical cases that involve propagation over long times and/or distances, a precise numerical calculation is very difficult or impossible, especially if many calculations are necessary in order to design, analyze or optimize the performance of the system under various conditions. The present work offers an efficient tool that can cope with such tasks and provide reliable results. In addition,
this work includes numerous analytical results that provide insight and understanding of the underlying physics, which would not be apparent from complex numerical calculations.
 
The framework offered in this work is quite simple. It requires calculation of only three static parameters: the initial wave-packet sizes $\sigma_j$ along the three Cartesian axes of the trap [Eq.~(\ref{eq:selfcons})], and three dynamic parameters: the three scaling factors $\lambda_j$ for the wave-packet along these axes [Eq.~(\ref{eq:ddlambda})]. This set of parameters, together with the center coordinates of the wave-packet, constitute 
all the wave-packet properties necessary for calculating the physical properties of each interferometer arm. Here we both demonstrate how these parameters are calculated and the resulting effects on interferometric performance. 
 
The theory of coherence of a two-state interferometer in Secs.~\ref{sec:spincoherence} generalizes the recent theory of Ref.~\cite{Roura2014} to include the whole range of atom-atom interactions. The phase diffusion theory in Sec.~\ref{sec:phasediffusion}
generalizes extensive previous work (see the beginning of Sec.~\ref{sec:phasediffusion}) concerning phase diffusion in a double-well or harmonic potential to the regime where the two interferometer arms carry wave-packets that dynamically evolve in space. In this case we find a novel contribution from the number dependence of the wave-packet expansion rate that may be of importance in some circumstances.

Finally, specific calculations presented in this paper for models of trapping and interferometric sequences are intended 
to demonstrate the basic physics and potential utility of the wave-packet evolution theory. Further work that uses this theory for analyzing previously published experimental results deserve separate publications. The present theory  
 (or unpublished versions) has already been used for analyzing experiments with Stern-Gerlach interferometers~\cite{Margalit2015,Margalit2019}, but it could also be implemented for analyzing experiments of other research groups (see, for example, Refs.~\cite{ Garcia2006,Burke2008,Ilo-Okeke2010,Fallen2015}). 
In addition, the theory presented here can be used for feasibility studies of future interferometric schemes that combine elements from both Secs.~\ref{sec:spincoherence} and~\ref{sec:phasediffusion} that were not explicitly discussed here, such as two-state interferometry with moving traps~\cite{Stevenson2015}, where 
effects of incomplete overlap at the output port, as well as phase diffusion, may be crucial to the interferometer performance. 
 
Finally, let us mention three possible extensions of the wave-packet evolution theory, beyond the scope of this paper, that would make it more general and effective. First, the current theory is based on the assumption that a quadratic expansion of the external potential around the wave-packet center is sufficient to describe the evolution. One would like to define quantitatively the range of validity of this assumption and examine the possible effects of higher-order terms of the potential. Second, we have not considered rotational effects when the axes of the time-dependent external potential do not coincide with the axes of the initial trap. This case could possibly  be treated in a way similar to what was presented in Ref.~\cite{Meister2017} and one would expect a synthesis of that method with the present work. Third, we have not provided an explicit form for the wave-packet envelope, which was assumed to be an implicit interpolation between a Gaussian and an inverted parabola. A more explicit approximation for the envelope in the initial trap and its evolution could possibly be worked out as an extension of the present work and provide more details regarding properties of the wave-packet that were not discussed here.

\acknowledgements
 
I am grateful to the members of the BGU atomchip group for useful discussions and helpful comments, particularly to Mark Keil, Yair Margalit, David Groswasser, Samuel Moukouri and Ron Folman. 
This work is funded in part by the Israel
Science Foundation (grant No. 856/18) and the German-Israeli DIP project (Hybrid devices: FO 703/2-1) supported
by the DFG. We also acknowledge support from
the Israeli Council for Higher Education.
 
\appendix
 
\section{Analytical expressions for BEC expansion in a waveguide}
\label{app:BECwaveguide}
 
To understand the oscillations of $\lambda_{\perp}$ in the waveguide potential, let us examine Eq.~(\ref{eq:ddlambda}) in the case where evolution along the longitudinal axis is much slower than the evolution along the radial axis. In this case the equations of motion for $\lambda_{\perp}$ can be written as
\be \ddot{\lambda}_{\perp}=\left(\nu_{\perp}^2+\tilde{\omega}_{\perp}^2\frac{\eta}{\lambda_x}\right)\frac{1}{\lambda_{\perp}^3}-\omega_{\perp}^2\lambda_{\perp}, 
\label{eq:ddlambda_waveguide} \ee
where $\lambda_x(t)$ is assumed to vary on a time scale that is much longer than the time scale determined by the frequency $\omega_{\perp}$. In this case Eq.~(\ref{eq:ddlambda_waveguide}) is equivalent to the classical equation of motion for a massive particle in a potential $V(q)=a/q^2+\frac{1}{2}m\omega^2q^2$. This potential has a minimum at $q_0^4=2a/m\omega^2$ and the frequency at the bottom of the potential is $\omega_0^2=\left. \partial^2V/\partial q^2\right|_{q=q_0}/m=6a/mq_0^4+\omega^2=4\omega^2$. It follows that the oscillations of $\lambda_{\perp}$ have a frequency that is twice the trap frequency and their center is given by 
\be \bar{\lambda}_{\perp}=\left[\left(\frac{(\nu_{\perp}}{\omega_{\perp}}\right)^2+\left(\frac{\tilde{\omega}_{\perp}}{\omega_{\perp}}\right)^2\frac{\eta}{\lambda_x}\right]^{1/4}, 
\label{eq:barlambda} \ee
such that $\bar{\lambda}_{\perp}\approx (\eta/\lambda_x)^{1/4}$ in the TF approximation (if the waveguide frequency is the same as the initial trapping frequency). 
In this last case the equation for the longitudinal scaling becomes $\ddot{\lambda}_x\approx \eta\omega_x(0)^2/\bar{\lambda}_{\perp}^2\lambda_x^2=\sqrt{\eta}\omega_x(0)^2/\lambda_x^{3/2}$. By analogy to a classical mass in a potential $V(q)=2a/\sqrt{q}$, where $a=2\sqrt{\eta}\omega_x(0)^2$, we find that after a long time $t\gg \omega_x(0)^{-1}$ the longitudinal cloud size expands with a constant rate $\dot{\lambda}_x=2\eta^{1/4}\omega_x(0)$. The shrinking of the cloud size in the transverse direction continues until it reaches the minimal uncertainty limit $\sigma_{\perp}(t)=\sigma_{\perp}(0)\bar{\lambda}_{\perp}\to \ell_{\perp}=\sqrt{\hbar/2m\omega_{\perp}}$, as can be verified from Eq.~(\ref{eq:barlambda}). 
 
\section{Derivation of the number-dependent phase}
\label{app:number_phase}
 
For estimating the overlap integral in Eq.~(\ref{eq:ovlp}) we use Eq.~(\ref{eq:Ansatz}) for the evolution of the wave-packet functions and a Gaussian approximation for the initial wave-packet $\Phi_0({\bf r})\propto \exp[-\sum_j r_j^2/4\sigma_j^2]$. We can then separate the variables $r_j=x,y,z$, so that 
\be A_a^{(n)}= e^{-i(n-1)(\varphi_a^{(n)}-\varphi_a^{(n-1)})}\prod_j {\cal A}_{a,j}^{(n)}, \ee
where $\varphi_a$ are the coordinate independent phases and 
\be {\cal A}_{a,j}^{(n)}\equiv C_{n,j}\int dx\, e^{-b_{n,j}x^2}, \ee
such that we omitted the arm index $a=L,R$ for simplicity, $C_{j,n}=(2\pi\lambda_{j,n}\lambda_{j,n-1}\sigma_j^2)^{-1/2}$ is a normalization constant, and
\be b_{n,j}=\frac{1}{4\sigma_j^2}\left(\frac{1}{\lambda_{j,n}^2}+\frac{1}{\lambda_{j,n-1}^2}\right) 
+\frac12 i(\alpha_j^{(n)}-\alpha_j^{(n-1)}). \ee
The single-coordinate integral yields
\be {\cal A}_{aj}^{(n)}=\left[\frac12(\xi_{j,n}+\xi_{j,n}^{-1}) +2i\epsilon_{j,n}\right]^{-(n-1)/2}\approx e^{-i\epsilon_{j,n}(n-1)}, 
\label{eq:Aajn} \ee
where $\xi_{j,n}\equiv \lambda_{j,n-1}/\lambda_{j,n}$ and 
\begin{eqnarray} \epsilon_{j,n} &=& \frac12 \lambda_{j,n}\lambda_{j,n-1}\sigma_j^2(\alpha_j^{(n)}-\alpha_j^{(n-1)}) \nonumber \\
&=& \frac{\lambda_{j,n}\lambda_{j,n-1}}{4\nu_j}\frac{\partial}{\partial n}\left(\frac{\dot{\lambda}_{j,n}}{\lambda_{j,n}}\right).
\label{eq:epsilon} \end{eqnarray} 
The last step in Eq.~(\ref{eq:Aajn}), where we approximate the absolute value to be 1, $|{\cal A}_{a,j}^{(n)}|=1$, is based on the following argument: as $\lambda_{j,n}(t)$ are determined by the solution of Eq.~(\ref{eq:ddlambda}) with $\eta_n=n/N$, it is expected that the difference $\lambda_{j,n}-\lambda_{j,n-1}$ scales like $eta_n-\eta_{n-1}=1/N$. It follows that $\frac12(\xi_{j,n}+\xi_{j,n}^{-1})\approx 1+O(1/N^2)$ and as $\alpha_j^{(n)}-\alpha_j^{(n-1)}$ is expected to scale like $1/N$ it also follows that its effect on the absolute value of ${\cal A}_{a,j}^{(n)}$ is proportional to $\epsilon_{j,n}^2=O(1/N^2)$. Hence even when the single-coordinate overlap integral is taken to the power of $(n-1)/2$ still the deviation from $|{\cal A}_{a,j}^{(n)}|=1$ is of the order $1/N$ and can be neglected for large $N$. 
Exceptions could happen in some extreme cases, for example, if $\lambda_{j,n}$ oscillates with a number-dependent frequency for a long time, such that the scaling factors for $n$ and $n-1$ become very different. Here we will not concentrate on such cases. 
We then have
\be A_a^{(n)}\approx e^{i(n-1)\epsilon_{j,a,n}}
e^{-i(n-1)(\varphi_a^{(n)}-\varphi_a^{(n-1)})}, 
\label{eq:Aan_1} \ee
where $\epsilon_{j,a,n}$ is given in Eq.~(\ref{eq:epsilon}) and the second term represents the difference between the global phases of the $n-1$ wave functions of $n$ and $n-1$ particles at the same arm. 
 
By collecting the global phases from all the wave functions appearing in the two configurations $k,N-k$ and $k-1,N-k+1$ we obtain
\be \chi_{k,N-k}-\chi_{k-1,N-k+1}=\left.\frac{\partial\chi_L^{(n)}}{\partial n}\right|_{k-\frac12}-\left. \frac{\partial\chi_R^{(n)}}{\partial n}\right|_{N-k+\frac12}, 
\ee
where $\chi_L^{(n)}$ and $\chi_R^{(n)}$ are integrals over time of Eq.~(\ref{eq:dotchi_an}). 
Therefore it follows that the sum of the global phases and the phases that come out of $A_L^{(k)}$ and $A_R^{(N-k+1)}$ in Eq.~(\ref{eq:Aan_1}) sum up to 
\be \delta\varphi_{k,k-1}=\tilde{\varphi}_L^{(k-\frac12)}-\tilde{\varphi}_R^{(N-k+\frac12)}, \ee
where
\be \tilde{\varphi}_a^{(n)}=
\frac{\partial\chi_a^{(n)}}{\partial n}+\sum_{j=1}^3 (n-1)\frac{\lambda_{j,a,n-\frac122}}{4\nu_j}\frac{\partial}{\partial n}\left(\frac{\dot{\lambda}_{j,a,n}}{\lambda_{j,a,n}}\right). \ee

Let us also note that under the same considerations that led to the approximation above, where the absolute value of the overlap integral was shown to be 1, we can also assume that the absolute value of the wave functions $|\Phi_a^{(n)}({\bf r},t)=\Phi_0(x/\lambda_1,y/\lambda_2,z/\lambda_3)$ can be assumed to be independent of the particle number $n$. It follows that the number dependence of the scaling factors $\lambda_{j,n}$ influences the phase but may be neglected when considering the wave function shape.

\end{document}